\pdfoutput = 1
\documentclass[12pt,a4paper]{article}

\usepackage[top=1in, bottom=1.25in, left=1in, right=1in]{geometry}

\usepackage{tikz}
\usetikzlibrary{shapes.geometric, arrows.meta, positioning}

\tikzstyle{box} = [rectangle, rounded corners, minimum width=28mm, minimum height=8mm,text centered, draw=black, fill=white]
\tikzstyle{obs} = [rectangle, rounded corners, minimum width=3mm, minimum height=3mm,text centered, draw=black, fill=red!20]
\tikzstyle{arrow} = [red,line width=1.5pt,-{Stealth[length=3mm]},shorten >=3pt]

\usepackage{amsmath}
\usepackage{graphicx}
\usepackage{xcolor}
\usepackage{booktabs}
\usepackage{rotating}
\usepackage[colorlinks=true, allcolors=blue]{hyperref}

\makeatletter
\g@addto@macro\bfseries{\boldmath}
\makeatother

\usepackage{ifthen}
\newboolean{articletitles}
\setboolean{articletitles}{true}
\newboolean{inbibliography}
\setboolean{inbibliography}{false}
\usepackage{cite}
\usepackage{mciteplus}

\begin{document}
%
%
%
\begin{titlepage}
\vspace*{-0.7truecm}
\begin{flushright}
Nikhef-2025-007 \\
SI-HEP-2025-09 \\
P3H-25-031
\end{flushright}

\vspace{1.4truecm}

\begin{center}
\renewcommand{\baselinestretch}{1.25}
\Large \textbf{How to tame penguins: Advancing to high-precision measurements of $\phi_d$ and $\phi_s$}
\end{center}

\vspace{0.8truecm}

\begin{center}
{\bf Kristof De Bruyn\,${}^{a,b}$, Robert Fleischer\,${}^{a,c}$, and Eleftheria Malami\,${}^{d}$}

\vspace{0.5truecm}

${}^a${\sl Nikhef, Science Park 105, 1098 XG Amsterdam, Netherlands}

${}^b${\sl Van Swinderen Institute for Particle Physics and Gravity, University of Groningen, 9747 Groningen, Netherlands}

${}^c${\sl  Faculty of Science, Vrije Universiteit Amsterdam,\\
1081 HV Amsterdam, Netherlands}

${}^d${\sl Center for Particle Physics Siegen, University of Siegen,\\
D-57068 Siegen, Germany}
\end{center}

\vfill

\begin{abstract}
The complex phases $\phi_d$ and $\phi_s$, associated with the mixing between neutral $B_q^0$ and $\bar B_q^0$ mesons ($q\in\{d,s\}$), are key observables to test the Standard Model and search for contributions from new physics.
They are conventionally determined from the measurements of mixing-induced CP violation in the decays $B_d^0\to J/\psi K^0$, $B_s^0\to J/\psi\phi$ and $B_s^0\to D_s^+D_s^-$.
To reach the highest possible precision on $\phi_d$ and $\phi_s$, it is crucial that corrections from next-to-leading order effects --- primarily associated with penguin decay topologies --- are accounted for.
The strategy adopted in this paper uses the $SU(3)$ flavour symmetry of QCD to relate the unknown contributions from penguin topologies to their counterparts in suitably-chosen control modes, where their effects are enhanced.
Utilising new CP asymmetry measurements from LHCb on the decays $B_s^0\to D_s^+D_s^-$, $B_d^0\to D^+D^-$, $B^+\to J/\psi K^+$ and $B^+\to J/\psi\pi^+$, as well as from Belle-II on the decay $B_d^0\to J/\psi\pi^0$, we present the current state-of-the-art picture on controlling the penguin contributions and extract $\phi_d$ and $\phi_s$ from the corresponding observables.
We explore the prospects for the end of the Belle-II and HL-LHC flavour physics programmes, and demonstrate the importance of measuring the control modes with future data.
\end{abstract}

\vfill

\noindent
January 2026

\end{titlepage}

\thispagestyle{empty}
~
\newpage

\setcounter{page}{1}

\section{Introduction}
The determination of the complex phases $\phi_d$ and $\phi_s$ associated with the mixing between neutral $B_q^0$ and $\bar B_q^0$ mesons, where $q\in\{d,s\}$, are key objectives \cite{LHCb:2018roe,Belle-II:2018jsg,ATLAS:2025lrr} for the particle physics experiments at the Large Hadron Collider (LHC) and SuperKEKB accelerator.
In the Standard Model (SM), these two phases are parametrised as
\begin{equation}\label{eq:phi_SM}
    \phi_d^{\text{SM}} = 2\beta\:, \qquad
    \phi_s^{\text{SM}} = 2\lambda^2\eta\:,
\end{equation}
where $\beta$ is one of the angles of the Unitarity Triangle (UT), and $\lambda$ and $\eta$ are two of the Wolfenstein parameters \cite{Wolfenstein:1983yz,Buras:1994ec} of the Cabibbo--Kobayashi--Maskawa (CKM) quark-mixing matrix \cite{Cabibbo:1963yz,Kobayashi:1973fv}.
Because of these simple dependences on SM parameters, both $\phi_d$ and $\phi_s$ provide interesting tests of the SM flavour sector, and thus allow us to search for indirect evidence of beyond the Standard Model (BSM) physics.
With Eq.\ \eqref{eq:phi_SM}, $\phi_s^{\text{SM}}$ can be predicted with a precision that is at least an order of magnitude smaller \cite{Barel:2020jvf,Charles:2015gya} than the current experimental world average \cite{HFLAV:2022pwe}, leaving ample room and motivation to improve the measurements of $\phi_s$.
For $\phi_d^{\text{SM}}$ the situation is less clear-cut \cite{DeBruyn:2022zhw} due to the unresolved tension between the inclusive and exclusive determinations of the CKM elements $|V_{ub}|$ and $|V_{cb}|$, which impacts the prediction for the apex of the UT.
Nonetheless, high precision measurements of both $\phi_d$ and $\phi_s$ remain a priority for Belle-II and the LHC experiments as a unique tool to test the CKM paradigm.

Experimentally, the phase $\phi_q$ is measured in Charge-Parity (CP) asymmetries generated by the interference between the $B_q^0$--$\bar B_q^0$ mixing process and the subsequent decay of the $B_q^0$ or $\bar B_q^0$ meson.
The phases $\phi_d$ and $\phi_s$ have been measured in a variety of decay channels, but the most precise measurements come from the ``golden modes" $B_d^0\to J/\psi K^0$ and $B_s^0\to J/\psi\phi$, respectively.
Throughout this paper, we use the label $B_d^0\to J/\psi K^0$ as shorthand notation to refer to both $B_d^0\to J/\psi K_{\text{S}}^0$ and $B_d^0\to J/\psi K_{\text{L}}^0$, and assume that CP violation in the kaon system has a negligible impact on the outcome of our analyses.
The two ``golden" decay channels have large branching fractions, which is favourable for the experimental measurement of their CP asymmetries, and their decay process is dominated by the colour-suppressed tree topology, simplifying the theoretical interpretation of these measurements.
If only the leading-order tree topologies are considered, then there is no direct CP asymmetry in the decay, and the resulting mixing-induced CP asymmetry is directly proportional to the mixing phase $\phi_q$.
The same reasoning can also be applied to the decay $B_s^0\to D_s^+D_s^-$, which is dominated by the colour-favoured tree topology, and provides an interesting cross-check for the measurement of $\phi_s$ in $B_s^0\to J/\psi\phi$.
However, $B_d^0\to J/\psi K^0$, $B_s^0\to J/\psi\phi$ and $B_s^0\to D_s^+D_s^-$ can also proceed via other decay topologies, which introduce small corrections to the leading-order result.
The largest corrections are due to penguin topologies, although $B_s^0\to J/\psi\phi$ and $B_s^0\to D_s^+D_s^-$ also receive even smaller contributions from exchange and penguin-annihilation topologies.
Given the current experimental precision and prospects for the end of the Belle-II and High Luminosity LHC (HL-LHC) programmes \cite{LHCb:2018roe,Belle-II:2018jsg,ATLAS:2025lrr}, these subleading contributions need to be accounted for if we wish to continue using $\phi_d$ and $\phi_s$ as a tool to search for BSM physics.
Including subleading effects, the CP asymmetries in $B_d^0\to J/\psi K^0$, $B_s^0\to J/\psi\phi$ and $B_s^0\to D_s^+D_s^-$ give access to effective mixing phases of the following structure:
\begin{equation}\label{eq:eff_mix_phase}
    \phi_{q,f}^{\text{eff}} \equiv \phi_q + \Delta\phi_q^f = \phi_q^{\text{SM}} + \phi_q^{\text{NP}} + \Delta\phi_q^f\:,
\end{equation}
where $f$ identifies the decay channel in which $\phi_{q,f}^{\text{eff}}$ was measured, $\Delta\phi_q^f$ is a decay-channel-specific hadronic phase shift due to the subleading SM processes, and $\phi_q^{\text{NP}}$ is the potential new physics (NP) contribution from BSM physics.
Eq.\ \eqref{eq:eff_mix_phase} thus illustrates that subleading SM processes can hide or mimic --- depending on the sign --- potential NP effects if not properly taken into account.
Accurate knowledge on the phase shift $\Delta\phi_q^f$ is thus crucial to complement the high-precision measurements of $\phi_{q,f}^{\text{eff}}$ and obtain the smallest possible uncertainty on $\phi_q$.
The determination of $\phi_q$ forms the main objective of this paper, while the interpretation of $\phi_q$ in terms of its SM and BSM contributions, and the subtleties associated with such an interpretation, have been discussed in Ref.\ \cite{DeBruyn:2022zhw}.

Strategies to calculate or constrain $\Delta\phi_q^f$ have been proposed for more than 25 years  \cite{Fleischer:1999nz,Fleischer:1999zi,Fleischer:1999sj,Ciuchini:2005mg,Faller:2008zc,Faller:2008gt,DeBruyn:2010hh,Ciuchini:2011kd,Jung:2012mp,Liu:2013nea,Jung:2014jfa,DeBruyn:2014oga,Frings:2015eva,Bel:2015wha,Barel:2020jvf,Davies:2023arm}.
In this paper, our main approach to control the impact of the penguin topologies relies on the $SU(3)$ flavour symmetry of QCD, and follows the original strategy of Refs.\ \cite{Fleischer:1999nz,Fleischer:1999zi}.
This $SU(3)$-based approach relates the decay amplitudes of $B_d^0\to J/\psi K^0$, $B_s^0\to J/\psi\phi$ and $B_s^0\to D_s^+D_s^-$ to those of control modes (see Section \ref{sec:control}) in which the penguin topologies are enhanced compared to the leading tree topology.
This new analysis is motivated by the recently updated measurement of the effective mixing phase $\phi_s^{\text{eff}}$ in $B_s^0\to D_s^+D_s^-$ and the CP asymmetries in $B_d^0\to D^+D^-$ by LHCb \cite{LHCb:2024gkk}, the new measurement of the CP asymmetries in $B_d^0\to J/\psi\pi^0$ by Belle-II \cite{Belle-II:2024hqw}, and the new measurement of the direct CP asymmetry difference between $B^+\to J/\psi \pi^+$ and $B^+\to J/\psi K^+$ by LHCb \cite{LHCb:2024exp}.

In this paper, we determine the mixing phases $\phi_d$ and $\phi_s$ from the CP asymmetry measurements in $B_d^0\to J/\psi K^0$, $B_s^0\to J/\psi\phi$ and $B_s^0\to D_s^+D_s^-$ while taking into account the hadronic phase shift due to the penguin topologies.
The analysis framework is based on previous work presented in Ref.\ \cite{Barel:2020jvf} for $B_d^0\to J/\psi K^0$ and $B_s^0\to J/\psi\phi$, and in Ref.\ \cite{Bel:2015wha} for $B_s^0\to D_s^+D_s^-$.
For the first time, we perform a simultaneous analysis of all three decay channels and their control modes, which are linked to each other due to their dependence on $\phi_d$ and $\phi_s$.
After presenting the picture emerging from the current data, we use our analysis framework to highlight the importance of measuring the control modes with the best possible precision at Belle-II and the HL-LHC.
The current experimental results show that the penguin contributions can be controlled.
However, they will become the dominant systematic uncertainty in the determination of $\phi_d$ if the control modes do not get the same attention as the golden modes in the Belle-II and HL-LHC flavour physics programmes.

The paper is organised as follows:
We discuss the available control modes for our strategy in Section \ref{sec:control}.
We summarise the theoretical framework in Section \ref{sec:framework}, and give an overview of the considered fit strategies in Section \ref{sec:fits}.
The experimental measurements which serve as inputs to our analysis are discussed in Section \ref{sec:inputs}.
The different fit strategies are presented in Section \ref{sec:current}.
Next, we explore the impact of $SU(3)$-symmetry breaking effects in Section \ref{sec:SU3}, which is a key limitation of the theoretical precision.
Then, in Section \ref{sec:future}, we illustrate how the picture could look like at the end of LHC Run 3 in summer 2026, and at the end of the Belle-II and HL-LHC programmes.
Finally, we conclude in Section \ref{sec:conclusion}.

\section{Control Modes}\label{sec:control}

\paragraph{$B_d^0\to J/\psi K_{\text{S}}^0$:}
For the decay $B_d^0\to J/\psi K_{\text{S}}^0$, the ideal control mode is the decay $B_s^0\to J/\psi K_{\text{S}}^0$ \cite{Fleischer:1999nz,DeBruyn:2014oga}.
These two decays are partnered through the $U$-spin symmetry and have a one-to-one relation between all their decay topologies, obtained by exchanging all down and strange quarks with one another.
When high-precision measurements of the $B_s^0\to J/\psi K_{\text{S}}^0$ CP asymmetries become available, relying solely on the decay $B_s^0\to J/\psi K_{\text{S}}^0$ to constrain the contributions from penguin topologies will minimise the theoretical uncertainties associated with our proposed method.
Long term, this is therefore the preferred strategy.

Given the currently available experimental precision on the $B_s^0\to J/\psi K_{\text{S}}^0$ CP asymmetries, it is advantageous to consider other control modes as well.
Here, the decay $B_d^0\to J/\psi\pi^0$ plays an important role \cite{Ciuchini:2005mg,Faller:2008zc}.
The decay $B_d^0\to J/\psi\pi^0$ is a general $SU(3)$ partner of $B_d^0\to J/\psi K^0$, and receives contributions from exchange and penguin-annihilation topologies, which have no counterpart in $B_d^0\to J/\psi K^0$.
These additional contributions introduce subleading theoretical uncertainties, which in the future will disfavour it compared to $B_s^0\to J/\psi K_{\text{S}}^0$.
However, with the currently available experimental data and precision, the impact of these additional topologies cannot yet be meaningfully quantified.
In the analysis presented here we therefore assume them to be negligible.
This assumption will need to be revisited when more precise experimental data becomes available.

Other general $SU(3)$ partners of $B_d^0\to J/\psi K^0$ include the $U$-spin pair $B^+\to J/\psi\pi^+$ and $B^+\to J/\psi K^+$, for which high precision CP asymmetry measurements are available \cite{LHCb:2024exp}.
They receive contributions from annihilation topologies, which have no counterpart in $B_d^0\to J/\psi K^0$, and thus also introduce additional subleading theoretical uncertainties.
Moreover, these decays are only sensitive to direct CP asymmetry, which provides a single constraint and therefore in a stand-alone analysis is insufficient to determine the penguin contributions.
These two decays thus always have to be used in combination with information from other decay channels.
For that reason, we only consider them here in the extension to our nominal fit scenario.

\paragraph{$B_s^0\to J/\psi\phi$:}
The strategy for controlling the impact of penguin topologies in the decay $B_s^0\to J/\psi\phi$ is more involved compared to $B_d^0\to J/\psi K^0$.
Firstly, the decay $B_s^0\to J/\psi\phi$ has two vector mesons in the final state, and therefore proceeds through a mixture of CP eigenstates.
This requires an analysis of the angular distributions of the decay products to disentangle the CP-odd and CP-even components, and makes the hadronic phase shift $\Delta\phi_q^f$, and thus also $\phi_q^{\text{eff}}$, depend on the polarisation state.
To correct for the polarisation-dependent hadronic shifts, the $B_s^0\to J/\psi\phi$ CP asymmetries have to be measured for each polarisation state individually.
The control modes need to match this structure, and thus should also have two vector mesons in the final state.
Note that the currently available experimental results do not always provide polarisation-dependent measurements of the $B_s^0\to J/\psi\phi$ CP asymmetries.
In this paper we therefore restrict ourselves to a polarisation-independent analysis of the penguin effects in $B_s^0\to J/\psi\phi$.

Secondly, the $\phi$ meson requires a more careful analysis, as was discussed in detail in Refs.\ \cite{Faller:2008gt,Gronau:2008kk}.
To good approximation it can be described as a pure $|s\bar s\rangle$ state.
This has two consequences: 
Firstly, it can mix with the orthogonal $\omega = (|u\bar u\rangle+|d\bar d\rangle)/\sqrt{2}$ state, although the impact of this mixing was found to be negligible \cite{Faller:2008gt,Gronau:2008kk}.
We will therefore ignore it here.
Secondly, the pure $|s\bar s\rangle$ state decomposes into a $SU(3)$ octet and a $SU(3)$ singlet contribution.
Our preferred control modes discussed below only address the $SU(3)$ octet contribution, leaving the $SU(3)$ singlet unconstrained.
Studying the singlet contribution relies on the future availability of CP asymmetry measurements in the $B_d^0\to J/\psi\omega$ and $B_s^0\to J/\psi\omega$ decays.

Keeping these assumptions in mind, we have identified the decay $B_d^0\to J/\psi\rho^0$ \cite{Fleischer:1999zi}, which has the same set of decay topologies as $B_s^0\to J/\psi\phi$, as the best-suited control mode.
In the future, when the necessary experimental data becomes available from decays such as $B_d^0\to J/\psi\phi$, $B_d^0\to J/\psi\omega$ and $B_s^0\to J/\psi\omega$, our assumptions and approximations can be cross-checked and improved upon.
Working towards this goal, Belle-II recently reported the first observation of the decay $B_d^0\to J/\psi\omega$ \cite{Belle-II:2024fgp}.
The pattern emerging from the current data is consistent with that observed in $B_d^0\to J/\psi K^0$, suggesting that these assumptions are acceptable at the current level of experimental precision.

Another control mode is the decay $B_s^0\to J/\psi \bar K^{*0}$.
However, there is a mismatch in the decay topologies as the exchange and penguin-annihilation diagrams of $B_s^0\to J/\psi\phi$ have no counterpart in $B_s^0\to J/\psi \bar K^{*0}$, introducing additional subleading theoretical uncertainties.
Moreover, this decay is only sensitive to direct CP asymmetry, and thus always has to be used in combination with information from other decay channels.

\paragraph{$B_s^0\to D_s^+D_s^-$:}
The decay $B_s^0\to D_s^+D_s^-$ has two pseudoscalars in the final state, which is again a different structure compared to $B_d^0\to J/\psi K^0$ and $B_s^0\to J/\psi\phi$.
Its ideal control mode, matching this structure, is the decay $B_d^0\to D^+D^-$ \cite{Fleischer:1999nz,Bel:2015wha} due to $U$-spin symmetry and the one-to-one relation between their decay topologies.
Due to this relation, we need not explicitly neglect the contributions from exchange and penguin-annihilation topologies, as they affect both decays equally.
Although, to be consistent in notation with the analyses of the decays $B_d^0\to J/\psi K^0$ and $B_s^0\to J/\psi\phi$, we will not be considering them in this analysis.

\section{Nomenclature}\label{sec:framework}
Let us introduce the framework we have developed to determine the penguin contributions from the experimental data, describing it from the point of view of a neutral $B_q^0$ meson decaying into a CP eigenstate $f$.
The decays $B_d^0\to J/\psi K^0$, $B_s^0\to J/\psi K_{\text{S}}^0$, $B_d^0\to J/\psi\pi^0$, $B_s^0\to D_s^+D_s^-$ and $B_d^0\to D^+D^-$ all fall into this category.
Also the polarisation-independent analysis of the decays $B_s^0\to J/\psi\phi$ and $B_d^0\to J/\psi\rho^0$ can be described in the exact same way.
The transition amplitudes for the $B_q^0\to f$ decay can be expressed as \cite{Fleischer:1999nz,Fleischer:1999zi}:
\begin{align}
    A(B_q^0\to f) & \equiv \phantom{\eta_f}\mathcal{N}_f\left[1-b_f e^{i\rho_f}e^{+i\gamma}\right]\:, \label{eq:TransAmp} \\
    A(\bar B_q^0\to f) & \equiv \eta_f\mathcal{N}_f\left[1-b_f e^{i\rho_f}e^{-i\gamma}\right]\:,
\end{align}
where $\eta_f$ is the CP-eigenvalue of the final state $f$.
In these two expressions, $\mathcal{N}_f$ is a CP-conserving normalisation factor which is dominated by the contribution from the tree topology, while the second term inside the brackets gives the relative contribution of the penguin topologies with respect to this normalisation factor.
This relative contribution is parametrised by the size $b_f$, a CP-conserving strong phase difference $\rho_f$ and a CP-violating weak phase difference given by the UT angle $\gamma$.

The time-dependent CP asymmetry for the $B_q^0\to f$ decay is defined as \cite{Fleischer:1999nz,Fleischer:1999zi}:
\begin{align}
    a_{\text{CP}}(t) & \equiv
    \frac{|A(B_q^0(t)\to f)|^2-|A(\bar B_q^0(t)\to f)|^2}{|A(B_q^0(t)\to f)|^2+|A(\bar B_q^0(t)\to f)|^2} \\
    & = \frac{\mathcal{A}_{\text{CP}}^{\text{dir}}\cos(\Delta m_qt)+\mathcal{A}_{\text{CP}}^{\text{mix}}\sin(\Delta m_qt)}{\cosh(\Delta\Gamma_qt/2)+\mathcal{A}_{\Delta\Gamma}\sinh(\Delta\Gamma_qt/2)}\:,
\end{align} 
where $\Delta m_q\equiv m^{(q)}_{\text{H}}-m^{(q)}_{\text{L}}$ and $\Delta\Gamma_q\equiv \Gamma_{\text{L}}^{(q)}-\Gamma_{\text{H}}^{(q)}$ are the mass and decay width difference between the heavy and light mass eigenstates of the $B_q$-meson system, respectively.
The time-dependent CP asymmetry is parametrised by three observables: 
the mass eigenstate rate asymmetry $\mathcal{A}_{\Delta\Gamma}$, the direct CP asymmetry $\mathcal{A}_{\text{CP}}^{\text{dir}}$ and the mixing-induced CP asymmetry $\mathcal{A}_{\text{CP}}^{\text{mix}}$.
These three observables are not independent from one another, but satisfy the relation
\begin{equation}
    \left[\mathcal{A}_{\text{CP}}^{\text{dir}}(B_q\to f)\right]^2 +
    \left[\mathcal{A}_{\text{CP}}^{\text{mix}}(B_q\to f)\right]^2  +
    \left[\mathcal{A}_{\Delta\Gamma}(B_q\to f)\right]^2 = 1\:.
\end{equation}
The CP asymmetries depend on the penguin parameters $b_f$ and $\rho_f$, and the $B_q^0$--$\bar B_q^0$ mixing phase $\phi_q$ as follows \cite{Fleischer:1999nz,Fleischer:1999zi}:
\begin{align}
    \mathcal{A}_{\text{CP}}^{\text{dir}}(B_q\to f) & = \frac{2 b_f \sin\rho_f\sin\gamma}{1-2b_f\cos\rho_f\cos\gamma+b_f^2}\:, \label{eq:Adir}\\
    \eta_f\mathcal{A}_{\text{CP}}^{\text{mix}}(B_q\to f) & = \left[ \frac{\sin\phi_q-2 b_f \cos\rho_f\sin(\phi_q+\gamma)+b_f^2\sin(\phi_q+2\gamma)}{1 - 2b_f\cos\rho_f\cos\gamma+b_f^2}\right]\:.\label{eq:Amix}
\end{align}
When only considering contributions from the leading order tree topologies and neglecting the doubly Cabibbo-suppressed penguin contributions, i.e.\ $b_f = 0$, these expressions simplify to the familiar forms 
\begin{equation}
    \mathcal{A}_{\text{CP}}^{\text{dir}} = 0\:,\qquad
    \eta_f\mathcal{A}_{\text{CP}}^{\text{mix}} = \sin\phi_q\:.
\end{equation}
Although both this paper and previous analyses \cite{Barel:2020jvf,Bel:2015wha} show that $b_f$ is small, both the current experimental uncertainties on the CP asymmetries in $B_d^0\to J/\psi K^0$, $B_s^0\to J/\psi\phi$, $B_s^0\to D_s^+D_s^-$ and their expected future precision no longer allow us to make the above approximation.
The determination on $\phi_d$ and $\phi_s$ through the CP asymmetries in Eqs.\ \eqref{eq:Adir} and \eqref{eq:Amix} thus requires knowledge on $b_f$ and $\rho_f$.

The $SU(3)$ flavour symmetry of QCD allows us to relate the hadronic parameters of the $\bar b\to \bar s c \bar c$ and $\bar b\to \bar d c \bar c$ transitions to one another, and thus to relate the normalisation factors and penguin contributions of the $B_d^0\to J/\psi K^0$, $B_s^0\to J/\psi\phi$, $B_s^0\to D_s^+D_s^-$ decays and their respective control channels.
Let us illustrate this for the decay $B_d^0\to J/\psi K_{\text{S}}^0$.
The expressions for the transition amplitudes and CP observables are obtained by substituting \cite{Fleischer:1999nz,Fleischer:1999zi}
\begin{equation}\label{eq:sub_ccs}
    \mathcal{N}_f \to \left(1-\frac{\lambda^2}{2}\right)\mathcal{A}'\:, \qquad
    b_f e^{i\rho_f} \to -\epsilon a' e^{i\theta'}
\end{equation}
where $\mathcal{A}'$ is a hadronic amplitude \cite{Barel:2020jvf} and 
\begin{equation}
    \epsilon \equiv \frac{\lambda^2}{1-\lambda^2} = 0.05237 \pm 0.00027\:.
\end{equation}
The numerical value is calculated using $\lambda \equiv |V_{us}|$ from Ref.\ \cite{Seng:2022wcw}.
$a'$ and $\theta'$ are the size and complex phase of a ratio between the hadronic amplitude of the penguin and tree decay topologies \cite{Barel:2020jvf}.
Together, $a'$ and $\theta'$ will be referred to as the penguin parameters.

The decay topologies in the control mode $B_s^0\to J/\psi K_{\text{S}}^0$ have a different dependence on the CKM matrix elements compared to $B_d^0\to J/\psi K_{\text{S}}^0$.
The equivalent substitutions to Eq.\ \eqref{eq:sub_ccs} are \cite{Fleischer:1999nz,Fleischer:1999zi}
\begin{equation}\label{eq:sub_ccd}
    \mathcal{N}_f \to -\lambda \mathcal{A}\:, \qquad
    b_f e^{i\rho_f} \to a e^{i\theta}\:.
\end{equation}
Comparing these to Eq.\ \eqref{eq:sub_ccs}, the contribution from the tree topology, i.e.\ the normalisation factor $\mathcal{N}_f$, is suppressed, while the relative contribution from the penguin topology, which no longer includes the factor $\epsilon$, is enhanced.
The former results in a small branching fraction for the $B_s^0\to J/\psi K_{\text{S}}^0$ decay, while the latter allows us to determine the penguin parameters $a$ and $\theta$ from the CP asymmetries.

Because there is a one-to-one correspondence between all the decay topologies in $B_d^0\to J/\psi K_{\text{S}}^0$ and $B_s^0\to J/\psi K_{\text{S}}^0$, the $U$-spin symmetry allows us to relate the penguin parameters to one another
\begin{equation}\label{eq:SU3_pen}
    a'e^{i\theta'} = ae^{i\theta}\:,
\end{equation}
as well as the hadronic amplitudes
\begin{equation}\label{eq:SU3_had}
    \mathcal{A}'=\mathcal{A}\:.
\end{equation}
Note that the CP asymmetries in Eqs.\ \eqref{eq:Adir} and \eqref{eq:Amix} do not depend on these hadronic amplitudes $\mathcal{A}^{(')}$, but only on the penguin parameters $a^{(')}$ and $\theta^{(')}$, which are ratios of amplitudes.
This is an important distinction for the theoretical uncertainties associated with the analysis strategy presented in this paper.
Eq.\ \eqref{eq:SU3_had} is susceptible to both factorisable and non-factorisable $SU(3)$-symmetry breaking.
Eq.\ \eqref{eq:SU3_pen}, on the other hand, only gets corrections from non-factorisable $SU(3)$-symmetry breaking as the leading factorisable $SU(3)$-symmetry breaking correction drops out in the ratio.
We will explore the impact of non-factorisable $SU(3)$-symmetry breaking in Section \ref{sec:SU3} by modifying Eq.\ \eqref{eq:SU3_pen} to allow for differences between $(a',\theta')$ and $(a,\theta)$.
The relation \eqref{eq:SU3_had} can be used to probe $SU(3)$ flavour symmetry, as has been done in Ref.\ \cite{Bel:2015wha,Fleischer:2010ca}.

Neglecting contributions from exchange and penguin-annihilation topologies, the above two relations can be extended to include the $B_d^0\to J/\psi \pi^0$ as a second control mode.
All three decays are examples of $B\to J/\psi +$\emph{Pseudoscalar} decays, and the $SU(3)$ flavour symmetry allows us to describe them using a single set of penguin parameters.
In order to distinguish these penguin parameters from the other decay channels analysed in this paper, we will label them as $a_{J/\psi P}$ and $\theta_{J/\psi P}$.

The decays $B_s^0\to J/\psi\phi$ and $B_d^0\to J/\psi\rho^0$ have two vector mesons in the final state, and are therefore expected to have different decay dynamics compared to the $B\to J/\psi +$ \emph{Pseudoscalar} decays.
Nonetheless, for the polarisation-independent analysis done in this paper the same formalism can be applied, and identical substitutions to Eqs.\ \eqref{eq:sub_ccs} and \eqref{eq:sub_ccd} can be written down.
To distinguish these $B\to J/\psi +$\emph{Vector} decays from the other two groups we will label the penguin parameters as $a_{J/\psi V}$ and $\theta_{J/\psi V}$.
We strongly encourage future experimental analyses of the $B\to J/\psi +$\emph{Vector} modes to provide polarisation-dependent measurements of the CP asymmetries as the hadronic phase shifts can depend on the polarisation of the final state.

Identical substitutions to Eqs.\ \eqref{eq:sub_ccs} and \eqref{eq:sub_ccd} can be written for the decay $B_s^0\to D_s^+D_s^-$ and its control mode $B_d^0\to D^+D^-$, respectively.
Different to the $B\to J/\psi +$\emph{Pseudoscalar} and $B\to J/\psi +$\emph{Vector} decays, these two $B\to DD$ decays have two pseudoscalars in the final state and proceed via colour-allowed instead of colour-suppressed tree topologies.
Because of these differences in the decay dynamics, the penguin parameters are expected to be different.
For the $B\to DD$ decays we will therefore label the penguin parameters as $a_{DD}$ and $\theta_{DD}$.

The extended control channels $B^+\to J/\psi\pi^+$, $B^+\to J/\psi K^+$ and $B_s^0\to J/\psi\bar K^{*0}$ do not decay into a CP eigenstate, and are therefore not subject to time-dependent CP violation. 
Nonetheless, they can still be described with the same formalism used for the other seven decays.
Their direct CP asymmetries are described by Eq.\ \eqref{eq:Adir}, but there is no mixing-induced CP violation and no counterpart to Eq.\ \eqref{eq:Amix}.
Neglecting contributions from annihilation topologies, identical substitutions to Eqs.\ \eqref{eq:sub_ccs} and \eqref{eq:sub_ccd} can be written down for the $U$-spin pair $B^+\to J/\psi\pi^+$ and $B^+\to J/\psi K^+$ in terms of the penguin parameters $a_{J/\psi P}$ and $\theta_{J/\psi P}$.
For $B_s^0\to J/\psi\bar K^{*0}$ an identical substitution to Eq.\ \eqref{eq:sub_ccs} can be written down in terms of the penguin parameters $a_{J/\psi V}$ and $\theta_{J/\psi V}$.

\section{Fit Strategies}\label{sec:fits}
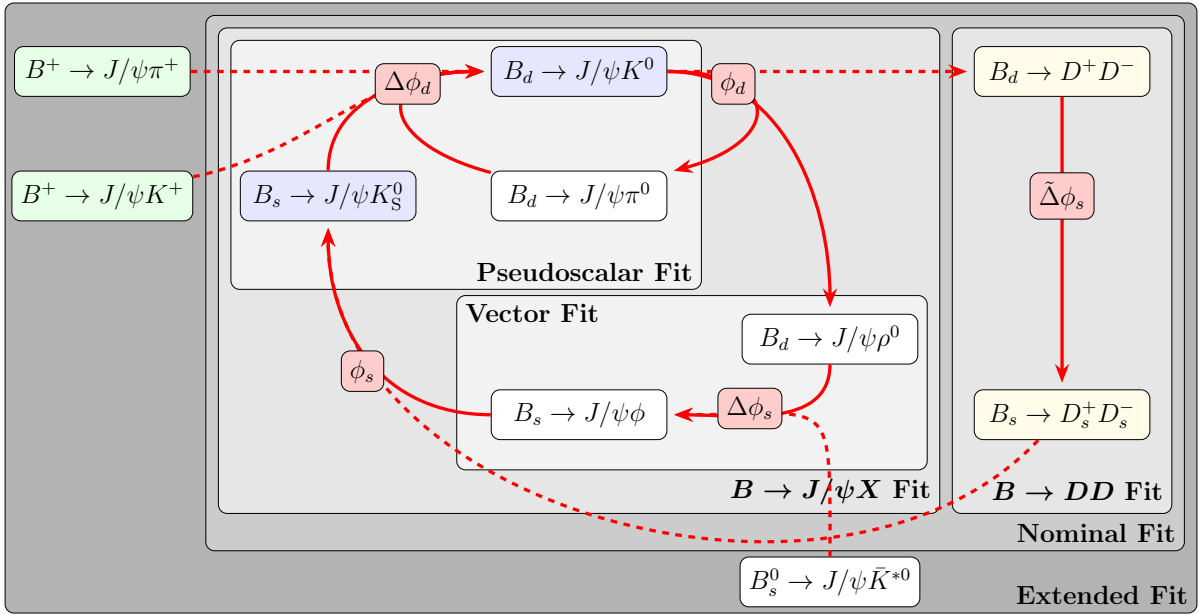
\begin{figure}
\resizebox{\textwidth}{!}{
\begin{tikzpicture}[node distance=20mm]
    \node (fitExt) [draw=black, fill=black!30, minimum width=190mm,minimum height=98mm, rounded corners]{};
    \node (fitAll) [below=2mm of fitExt.north east,anchor=north east, xshift=-2mm, draw=black, fill=black!20, minimum width=156mm,minimum height=86mm, rounded corners]{};
    \node (fitJX) [below=2mm of fitAll.north west, anchor=north west, xshift=2mm, draw=black, fill=black!10, minimum width=115mm,minimum height=78mm, rounded corners]{};
    \node (fitPS) [below=2mm of fitJX.north west,anchor=north west, xshift=2mm, draw=black, fill=black!5, minimum width=75mm,minimum height=40mm, rounded corners]{};
    \node (fitV) [above=7mm of fitJX.south east,anchor=south east, xshift=-2mm, draw=black, fill=black!5, minimum width=75mm,minimum height=28mm, rounded corners]{};
    
    \node (fitDD) [below=2mm of fitAll.north east, anchor=north east, xshift=-2mm, draw=black, fill=black!10, minimum width=35mm,minimum height=78mm, rounded corners]{};

    \node (textPS) [above=0mm of fitPS.south east, anchor=south east, font=\boldmath\bf] {Pseudoscalar Fit};
    \node (textV) [above=0mm of fitV.north west, anchor=north west, font=\boldmath\bf] {Vector Fit};
    \node (textJX) [above=0mm of fitJX.south east, anchor=south east, font=\boldmath\bf] {$B\to J/\psi X$ Fit};
    \node (textDD) [above=0mm of fitDD.south east, anchor=south east, font=\boldmath\bf] {$B\to DD\vphantom{\psi}$ Fit};
    \node (textAll) [above=0mm of fitAll.south east, anchor=south east, font=\boldmath\bf] {Nominal Fit};
    \node (textExt) [above=0mm of fitExt.south east, anchor=south east, font=\boldmath\bf] {Extended Fit};

    \node (BdJpsiK) [box, fill=blue!10, below=3mm of fitJX.north, anchor=north] {$B_d\to J/\psi K^0$};
    \node (BdJpsiPi) [box, below of=BdJpsiK] {$B_d\to J/\psi \pi^0$};
    \node (BsJpsiK) [box, fill=blue!10, left of=BdJpsiPi, xshift=-20mm] {$B_s\to J/\psi K_{\text{S}}^0$};
    \node (BdJpsiRho) [box, right of=BdJpsiPi, xshift=20mm, yshift=-23mm] {$B_d\to J/\psi \rho^0$};
    \node (BsJpsiPhi) [box, below=27mm of BdJpsiPi] {$B_s\to J/\psi \phi$};
    \node (BdDD) [box, fill=yellow!10, right of=BdJpsiK, xshift=57mm] {$B_d\to D^+D^-$};
    \node (BsDsDs) [box, fill=yellow!10, right of=BsJpsiPhi, xshift=57mm] {$B_s\to D_s^+D_s^-$};
    \node (BuJpsiK) [box, fill=green!10, left of=BsJpsiK, xshift=-16mm] {$B^+\to J/\psi K^+$};
    \node (BuJpsiPi) [box, fill=green!10, above of=BuJpsiK] {$B^+\to J/\psi \pi^+$};
    \node (BsJpsiKst) [box, below of=BdJpsiRho, yshift=-19mm] {$B_s^0\to J/\psi \bar K^{*0}$};

    \draw[arrow] (BdJpsiK) to [out=0,in=90] (BdJpsiRho);
    \draw[arrow] (BdJpsiK) to [out=0,in=15,looseness=3] (BdJpsiPi);
    \draw[arrow, dashed] (BdJpsiK) to [out=0,in=180] (BdDD);

    \draw[arrow] (BdJpsiRho) to [out=270,in=0] (BsJpsiPhi);
    \draw[arrow] (BdDD) to [out=270,in=90] (BsDsDs);

    \draw[arrow] (BsJpsiPhi) to [out=180,in=270, looseness=1.1] (BsJpsiK);
    \draw[arrow, dashed] (BsDsDs) to [out=227,in=270, looseness=1.] (BsJpsiK);

    \draw[arrow] (BsJpsiK) to [out=90,in=180] (BdJpsiK);
    \draw[arrow] (BdJpsiPi) to [out=165,in=180,looseness=3] (BdJpsiK);

    \draw[arrow, dashed] (BsJpsiKst) to [out=90,in=0, looseness=2] (BsJpsiPhi);
    \draw[arrow, dashed] (BuJpsiPi) to [out=0,in=180] (BdJpsiK);
    \draw[arrow, dashed] (BuJpsiK) to [out=12,in=180] (BdJpsiK);

    \node (phid) [obs, right=7mm of BdJpsiK, yshift=-2mm] {$\phi_d$};
    \node (phis) [obs, left=17mm of BsJpsiPhi, yshift=7mm] {$\phi_s$};
    \node (Dphid) [obs, left=8mm of BdJpsiK, yshift=-2mm] {$\Delta\phi_d$};
    \node (Dphis) [obs, right=8mm of BsJpsiPhi, yshift=1mm] {$\Delta\phi_s$};
    \node (Dphis2) [obs, below of=BdDD] {$\tilde\Delta\phi_s$};
\end{tikzpicture}
} 
    \caption{Schematic overview of the fit strategies considered in this paper, and the key quantities (highlighted in red) they can determine.
    The other colours highlight the $U$-spin partner decays.}
    \label{fig:models}
\end{figure}

In this paper we determine and compare the penguin parameters for six different fit strategies, which are illustrated in Fig.\ \ref{fig:models}.
These strategies are not all independent from one another, but combine more and more decay channels together to take into account the correlations between them.
This culminates in our nominal fit, which combines information from the $B_d^0\to J/\psi K_{\text{S}}^0$, $B_s^0\to J/\psi\phi$, $B_s^0\to D_s^+D_s^-$ decays and their primary control modes.
We have chosen to present the outcome of the intermediate and extended fit strategies as well to transparently show how the nominal fit has been built up and connects to previous analyses \cite{Faller:2008zc,DeBruyn:2014oga,Barel:2020jvf}.
When combining information from multiple decay channels there is a trade-off between increasing the experimental precision on the one hand and the theoretical assumptions made to obtain the result on the other.
Our assumptions are summarised at the end of this section.

\paragraph{Pseudoscalar Fit}
Using external inputs for the UT angle $\gamma$ and the mixing phase $\phi_s$, the direct and mixing-induced CP observables in $B_s^0\to J/\psi K_{\text{S}}^0$ can be used to determine the penguin parameters $a$ and $\theta$ \cite{Fleischer:1999nz}.
The decay topologies of $B_s^0\to J/\psi K_{\text{S}}^0$ can be mapped one-to-one to those of the $B_d^0\to J/\psi K^0$ by exchanging all down and strange quarks with one another.
These two decays are related via the $U$-spin symmetry, resulting in relation \eqref{eq:SU3_pen}. 
The solution for $a$ and $\theta$ therefore implies a solution for $a'$ and $\theta'$, allowing us to determine $\Delta\phi_d(a',\theta')$ and thus translate the measurement of $\phi_d^{\text{eff}}$ into a measurement of the mixing phase $\phi_d$ itself.
A similar strategy using $B_d^0\to J/\psi \pi^0$ was proposed in Refs.\ \cite{Ciuchini:2005mg,Faller:2008zc}.
Here there is no one-to-one mapping between all the decay topologies as $B_d^0\to J/\psi \pi^0$ also gets contributions from exchange and penguin-annihilation topologies, which have no counterpart in $B_d^0\to J/\psi K^0$.
However, as these topologies are expected to be even more suppressed than the penguin contributions, the $SU(3)$-symmetry strategy can still be employed and the impact of ignoring the additional decay topologies is assumed to be negligible.
When high precision measurements of the CP asymmetries in $B_s^0\to J/\psi K_{\text{S}}^0$ and $B_d^0\to J/\psi \pi^0$ become available, this assumption will have to be reconsidered.
At that stage, the usage of $B_d^0\to J/\psi \pi^0$ as a control mode will no longer be necessary, which minimises the theoretical uncertainty associated with the fit strategy.
However, given the current experimental picture, the usage of both control modes is still beneficial to constrain the penguin parameters $a_{J/\psi P}$ and $\theta_{J/\psi P}$, and thus also the hadronic shift $\Delta\phi_d$.
We will refer to this strategy, using both $B_s^0\to J/\psi K_{\text{S}}^0$ and $B_d^0\to J/\psi \pi^0$, as the ``Pseudoscalar" fit in our numerical comparison below.

\paragraph{Vector Fit}
Using external inputs for the UT angle $\gamma$ and the mixing phase $\phi_d$, the direct and mixing-induced CP observables in $B_d^0\to J/\psi \rho^0$ determine the penguin parameters $a_{J/\psi V}$ and $\theta_{J/\psi V}$ \cite{Fleischer:1999zi}.
Using the relation \eqref{eq:SU3_pen}, the solution for $a_{J/\psi V}$ and $\theta_{J/\psi V}$ determines the hadronic shift $\Delta\phi_s$ and thus allows us to determine $\phi_s$ from the measurements of $\phi_s^{\text{eff}}$ in $B_s^0\to J/\psi\phi$.
We will refer to this strategy as the ``Vector" fit in our numerical comparison below.

\paragraph{$B\to J/\psi X$ Fit}
The main limitation of the ``Pseudoscalar" and ``Vector" fit strategies is that they still require external inputs on one of the two mixing phases, which themselves need to be corrected for penguin effects, while aiming to determine the other of the two mixing phases in the presence of penguin topologies.
In Ref.\ \cite{Barel:2020jvf}, we therefore proposed to combine both fit strategies and simultaneously determine the mixing phases $\phi_d$ and $\phi_s$, as well as the pseudoscalar and vector penguin parameters.
This combined fit, which we will refer to as the ``$B\to J/\psi X$" fit, has the big advantage that it only requires external input on the UT angle $\gamma$.

\paragraph{$B\to DD$ Fit}
For the decay $B_s^0\to D_s^+D_s^-$ the situation is identical to the ``Pseudoscalar" and ``Vector" fit strategies.
Using external inputs for the UT angle $\gamma$ and the mixing phase $\phi_d$, the direct and mixing-induced CP observables in $B_d^0\to D^+D^-$ determine the penguin parameters $a_{DD}$ and $\theta_{DD}$.
The decay topologies of $B_d^0\to D^+D^-$ can be mapped one-to-one to those of the $B_s^0\to D_s^+D_s^-$ channel by exchanging all down and strange quarks with one another.
The two decays are thus also related via the $U$-spin symmetry.
Using the relation \eqref{eq:SU3_pen}, the solution for $a_{DD}$ and $\theta_{DD}$ determines the hadronic shift $\Delta\phi_s$ and thus allows us to determine $\phi_s$ from the measurements of $\phi_s^{\text{eff}}$ in $B_s^0\to D_s^+D_s^-$.
We will refer to this strategy, which was previously analysed in Ref.\ \cite{Bel:2015wha}, as the ``$B\to DD$" fit.
As a stand-alone analysis the ``$B\to DD$" fit has the same limitation as the ``Vector" fit.

\paragraph{Nominal Fit}
It is possible to combine the ``$B\to J/\psi X$" and ``$B\to DD$" fit strategies, allowing a simultaneous and self-consistent determination of $\phi_d$ and $\phi_s$ using all three groups of decay channels.
In this paper, we consider this to be our ``Nominal" fit strategy.

\paragraph{Extended Fit}
Finally, we can extend the ``Nominal" fit by including additional $SU(3)$ partners for the decays $B_d^0\to J/\psi K^0$ and $B_s^0\to J/\psi\phi$.
The decay $B^+\to J/\psi K^+$ is related to $B_d^0\to J/\psi K^0$ by exchanging the spectator $u$ and $d$ quarks with one another.
The decay $B^+\to J/\psi\pi^+$ is a general $SU(3)$ partner of $B_d^0\to J/\psi K^0$, and $U$-spin partner of $B^+\to J/\psi K^+$.
Both decays receive additional contributions from annihilation topologies that are not present in $B_d^0\to J/\psi K^0$.
The decay $B_s^0\to J/\psi\bar K^{*0}$ is a partner of $B_s^0\to J/\psi\phi$.
Contrary to $B_s^0\to J/\psi\phi$ it only gets contributions from tree and penguin topologies, and not from exchange and penguin-annihilation topologies.
All three additional $SU(3)$ partners are only sensitive to CP violation in decay, thus providing only one experimental constraint.
Individually, they therefore cannot be used to determine the penguin parameters.
However, when combined with the other control channels from the nominal fit, we can explore the impact these additional inputs have on the mixing phases and penguin parameters.
This final strategy is referred to as the ``Extended" fit.

\paragraph{Assumptions}
Given the current experimental precision on the measured CP asymmetries of the control modes, all fit strategies make the following three assumptions:
\begin{enumerate}
    \item We perform a polarisation-independent analysis of the $B\to J/\psi+$\emph{Vector} channels.
    \item We assume perfect $SU(3)$ flavour symmetry.
    \item We assume the contributions from annihilation, exchange, and penguin-annihilation topologies are subdominant to those from the penguin topologies and can therefore be excluded from our analyses.
\end{enumerate}

The $SU(3)$ flavour symmetry that allows us to relate the penguin contributions in $B_d^0\to J/\psi K^0$, $B_s^0\to J/\psi\phi$, $B_s^0\to D_s^+D_s^-$ and their control modes through the relations \eqref{eq:SU3_pen} and \eqref{eq:SU3_had} is only exact in the limit that the up, down and strange quarks have the same mass.
It is therefore expected that these relations are subject to $SU(3)$-symmetry breaking effects, which can be split in two categories: factorisable and non-factorisable contributions.
Eq.\ \eqref{eq:SU3_had} is subject to both, while Eq.\ \eqref{eq:SU3_pen} only gets contributions from non-factorisable effects.
Since this analysis does not make use of branching fraction information, relying exclusively on CP asymmetry observables, Eq.\ \eqref{eq:SU3_had} is not needed and factorisable $SU(3)$-breaking effects do not enter the analysis.
Instead, we are only left with the non-factorisable effects.
The nominal results provided in Section \ref{sec:current} assume the $SU(3)$ flavour symmetry to hold exactly, and thus do not account for possible non-factorisable $SU(3)$-breaking effects.
In Section \ref{sec:SU3} we explore the impact of non-factorisable $SU(3)$-breaking effects on the fit results and assign a systematic uncertainty due to the $SU(3)$-symmetry assumption on the values of $\phi_d$ and $\phi_s$.

The decays $B_d^0\to J/\psi\pi^0$, $B_s^0\to J/\psi\phi$, $B_d^0\to J/\psi\rho^0$, $B_s^0\to D_s^+D_s^-$ and $B_d^0\to D^+D^-$ get contributions from exchange and penguin-annihilation topologies in addition to the tree and penguin topologies accounted for in our analysis.
Likewise, the decays $B^+\to J/\psi\pi^+$ and $B^+\to J/\psi K^+$ also get contributions from annihilation topologies.
However, since these contributions are expected to be even smaller than those from the penguin topologies, we do not include them in the current analysis.
When higher precision measurements of the CP asymmetry observables become available in the future, this decision, as well as the choice of control modes to include in the analyses, will need to be revisited.

\section{Experimental Inputs}\label{sec:inputs}
The fit strategies use the following inputs:
\begin{itemize}
    \item \textbf{External constraint on the CKM element $|V_{us}| = \lambda$:}
    We use the experimental average
\begin{equation}
    |V_{us}|= 0.22308 \pm 0.00055\:.
\end{equation}
    from measurements in the $K\to\pi\ell\nu_{\ell}$ decay channels \cite{Seng:2022wcw}.
    \item \textbf{External constraint on the CKM angle $\gamma$:}
    There are three main strategies to determine the CKM angle $\gamma$: from time-independent measurements of $B\to DK$ decays \cite{LHCb:2016mag,HFLAV:2022pwe}, from a time-dependent analysis of the CP asymmetries in $B_s^0\to D_s^{\mp}K^{\pm}$ \cite{LHCb:2024xyw}, and from an isospin analysis of the $B\to\pi\pi,\rho\pi,\rho\rho$ decays \cite{Charles:2017evz,HFLAV:2022pwe}.
    However, not all three strategies are suitable for our analysis here.
    The time-dependent analysis of $B_s^0\to D_s^{\mp}K^{\pm}$ actually measures $\phi_s + \gamma$, and thus requires external input on $\phi_s$, which we aim to determine in this analysis, to extract $\gamma$.
    Likewise, the isospin analysis of the $B\to\pi\pi,\rho\pi,\rho\rho$ decays actually measures $\phi_d + 2\gamma$, and thus requires external input on $\phi_d$, which we aim to determine in this analysis, to extract $\gamma$.
    For the analysis in this paper, we therefore only make use of the time-independent measurements of $B\to DK$ decays to constrain $\gamma$.

    The average from time-independent $B\to DK$ decays, prepared by the Heavy Flavour Averaging Group (HFLAV) \cite{HFLAV:2022pwe} for the \emph{58th Rencontres de Moriond} conference and calculated by the authors using the GammaCombo framework \cite{LHCb:2016mag}, equals
\begin{equation}
    \gamma = (65.6_{-3.0}^{+2.9})^{\circ}\:.
\end{equation}
    \item \textbf{CP asymmetries of the decay $B_d^0\to J/\psi K^0$:}
    The decay $B_d^0\to J/\psi K^0$ is a CP-odd final state ($\eta_{\text{CP}}=-1$).
    Its CP asymmetries have been measured by the BaBar, Belle, LHCb and Belle-II experiments.
    The BaBar, Belle and LHCb ``Run 1" measurements are combined in the 2022 average from HFLAV \cite{HFLAV:2022pwe}
\begin{align}
    \mathcal{A}_{\text{dir}}^{\text{CP}}(B_d^0\to J/\psi K^0) & = -0.007 \pm 0.012 \text{(stat)} \pm 0.014 \text{(syst)}\:, \\
    \eta_{\text{CP}}\mathcal{A}_{\text{mix}}^{\text{CP}}(B_d^0\to J/\psi K^0) & = \phantom{-}0.690 \pm 0.017 \text{(stat)} \pm 0.006 \text{(syst)}\:.
    \label{eq:sin2beta_HFLAV}
\end{align}
    Belle-II published the measurement \cite{Belle-II:2024lwr}
\begin{align}
    \mathcal{A}_{\text{dir}}^{\text{CP}}(B_d^0\to J/\psi K^0) & = -0.035 \pm 0.026 \text{(stat)} \pm 0.029 \text{(syst)}\:, \\
    \eta_{\text{CP}}\mathcal{A}_{\text{mix}}^{\text{CP}}(B_d^0\to J/\psi K^0) & = \phantom{-}0.724 \pm 0.035 \text{(stat)} \pm 0.009 \text{(syst)}\:,
    \label{eq:sin2beta_Belle2}
\end{align}
    with a correlation $\rho=+0.09$, using a data sample corresponding to an integrated luminosity of $362\text{fb}^{-1}$.
    LHCb published a second measurement \cite{LHCb:2023zcp}
\begin{align}
    \mathcal{A}_{\text{dir}}^{\text{CP}}(B_d^0\to J/\psi K^0) & = \phantom{-}0.015 \pm 0.013 \text{(stat)} \pm 0.003 \text{(syst)}\:, \label{eq:Adir_Bd2JpsiK}\\
    \eta_{\text{CP}}\mathcal{A}_{\text{mix}}^{\text{CP}}(B_d^0\to J/\psi K^0) & = \phantom{-}0.722 \pm 0.014 \text{(stat)} \pm 0.007 \text{(syst)}\:,
    \label{eq:sin2beta_LHCb2}
\end{align}
    with a correlation $\rho=+0.437$, using the ``Run 2" data sample corresponding to an integrated luminosity of $6\text{fb}^{-1}$.
    We use these three measurements as independent inputs in our fit.

    The weighted average of the mixing-induced CP asymmetry measurements in Eqs.\ \eqref{eq:sin2beta_HFLAV}, \eqref{eq:sin2beta_Belle2} and \eqref{eq:sin2beta_LHCb2} corresponds to an effective mixing phase
\begin{equation}\label{eq:phid_eff}
    \phi_d^{\text{eff}} = \left(45.12 \pm 0.94\right)^{\circ}\:,
\end{equation}
    taking the correlations into account.
    \item \textbf{CP asymmetries of the decay $B_s^0\to J/\psi K_{\text{S}}^0$:}
    The decay $B_s^0\to J/\psi K_{\text{S}}^0$ is a CP-odd final state ($\eta_{\text{CP}}=-1$).
    We use the LHCb measurement \cite{LHCb:2015brj}:
\begin{align}
    \mathcal{A}_{\text{dir}}^{\text{CP}}(B_s^0\to J/\psi K_{\text{S}}^0) & = -0.28 \pm 0.41 \text{(stat)} \pm 0.08 \text{(syst)}\:, \label{eq:Adir_Bs2JpsiKs} \\
    \eta_{\text{CP}}\mathcal{A}_{\text{mix}}^{\text{CP}}(B_s^0\to J/\psi K_{\text{S}}^0) & = \phantom{-}0.08 \pm 0.40 \text{(stat)} \pm 0.08 \text{(syst)}\:, \label{eq:Amix_Bs2JpsiKs}
\end{align}
    taking into account the correlation $\rho=+0.06$ between both observables.
    This result is obtained using the ``Run 1" data sample, corresponding to an integrated luminosity of $3\text{fb}^{-1}$.
    \item \textbf{CP asymmetries of the decay $B_d^0\to J/\psi\pi^0$:}
    The decay $B_d^0\to J/\psi\pi^0$ is a CP-odd final state ($\eta_{\text{CP}}=-1$).
    Its CP asymmetries have been measured by the BaBar and Belle experiments using their full data sets.
    The average of these two results provided by HFLAV \cite{HFLAV:2022pwe} is
\begin{align}
    \mathcal{A}_{\text{dir}}^{\text{CP}}(B_d^0\to J/\psi\pi^0) & = 0.04 \pm 0.12\:, \label{eq:Adir_Bd2JpsiPi_B1}\\
    \eta_{\text{CP}}\mathcal{A}_{\text{mix}}^{\text{CP}}(B_d^0\to J/\psi\pi^0) & = 0.86 \pm 0.14\:, \label{eq:Amix_Bd2JpsiPi_B1}
\end{align}
    taking into account the correlation $\rho=-0.08$ between both observables.
    The Belle-II collaboration recently released a new measurement \cite{Belle-II:2024hqw}
\begin{align}
    \mathcal{A}_{\text{dir}}^{\text{CP}}(B_d^0\to J/\psi\pi^0) & = 0.13 \pm 0.12\text{(stat)} \pm 0.03 \text{(syst)}\:, \label{eq:Adir_Bd2JpsiPi_B2} \\
    \eta_{\text{CP}}\mathcal{A}_{\text{mix}}^{\text{CP}}(B_d^0\to J/\psi\pi^0) & = 0.88 \pm 0.17\text{(stat)} \pm 0.03 \text{(syst)}\:, \label{eq:Amix_Bd2JpsiPi_B2}
\end{align}
    taking into account the statistical correlation $\rho=+0.08$ between both observables.
    This result is obtained using a data sample corresponding to an integrated luminosity of 365 fb$^{-1}$.
    \item \textbf{CP asymmetries of the decay $B_s^0\to J/\psi\phi$:}
    Assuming a universal phase for all three polarisation states, the effective mixing phase $\phi_s^{\text{eff}}$ has been measured by the CDF, D0, ATLAS, CMS and LHCb collaborations.
    The HFLAV average \cite{HFLAV:2022pwe} for all $B_s^0\to J/\psi KK$ modes, which we interpret to correspond to the $B_s^0\to J/\psi\phi$ decay, is
\begin{equation}\label{eq:phis_eff}
    \phi_s^{\text{eff}} = -0.061 \pm 0.014 = (-3.50 \pm 0.80)^{\circ}\:.
\end{equation} 
    Recently, CMS has released a new measurement \cite{CMS:2024znt}
\begin{equation}
    \phi_s^{\text{eff}} = -0.073 \pm 0.023\text{(stat)} \pm 0.007 \text{(syst)}\:,
\end{equation}
    using a data sample corresponding to an integrated luminosity of $96.5\text{fb}^{-1}$ collected in 2017-2018.
    We combine both results together with $\lambda_{J/\psi\phi} = 0.995 \pm 0.009$ and $\eta_{\text{CP}}=-1$ to calculate the polarisation-independent CP asymmetries
\begin{align}
    \mathcal{A}_{\text{dir}}^{\text{CP}}(B_s^0\to J/\psi \phi) & = \phantom{-}0.005 \pm 0.009\:, \\
    \eta_{\text{CP}}\mathcal{A}_{\text{mix}}^{\text{CP}}(B_s^0\to J/\psi \phi) & = -0.061 \pm 0.014\:,
\end{align}
    which are used in the fit framework, similar to the other decay channels.
    \item \textbf{CP asymmetries of the decay $B_d^0\to J/\psi \rho^0$:}
    Because the $B_d^0\to J/\psi \rho^0$ has two vector particles in the final state, its CP asymmetries can depend on the polarisation of the final state.
    LHCb has made a polarisation-dependent measurement in Ref.\ \cite{LHCb:2014xpr} using the ``Run 1" data sample.
    However, for the $B_s^0\to J/\psi\phi$ decay the latest measurements only provide polarisation-independent results.
    To have a consistent treatment in our fit framework, we will therefore restrict our analysis to only use the polarisation-independent measurement from Ref.\ \cite{LHCb:2014xpr}
\begin{align}
    \mathcal{A}_{\text{dir}}^{\text{CP}}(B_d^0\to J/\psi \rho^0) & = -0.063 \pm 0.056 \text{(stat)} \pm 0.019 \text{(syst)}\:, \label{eq:Adir_Bd2JpsiRho}\\
    \eta_{\text{CP}}\mathcal{A}_{\text{mix}}^{\text{CP}}(B_d^0\to J/\psi \rho^0) & = \phantom{-}0.66\phantom{0} \pm 0.13\phantom{0} \text{(stat)} \pm 0.09\phantom{0} \text{(syst)}\:, \label{eq:Amix_Bd2JpsiRho}
\end{align}
    taking into account the correlation $\rho=+0.01$ between both observables, and $\eta_{\text{CP}}=-1$.
    \item \textbf{CP asymmetries of the decay $B_s^0\to D_s^+D_s^-$:}
    The decay $B_s^0\to D_s^+D_s^-$ is a CP-even final state ($\eta_{\text{CP}}=+1$).
    We use the LHCb combination \cite{LHCb:2024gkk} of their ``Run 1" and ``Run 2" results
\begin{align}
    \phi_s^{\text{eff}} & = -0.055 \pm 0.090\text{(stat)} \pm 0.021\text{(syst)}\:,\\
    \lambda_{D_sD_s} & = \phantom{-}1.054 \pm 0.099\text{(stat)} \pm 0.020\text{(syst)}
\end{align} 
    to calculate the CP asymmetries
\begin{align}
    \mathcal{A}_{\text{dir}}^{\text{CP}}(B_s^0\to D_s^+D_s^-) & = -0.053 \pm 0.096\:, \\
    \eta_{\text{CP}}\mathcal{A}_{\text{mix}}^{\text{CP}}(B_s^0\to D_s^+D_s^-) & = -0.055 \pm 0.092\:,
\end{align}
    which are used in the fit framework, similar to the other decay channels.
    \item \textbf{CP asymmetries of the decay $B_d^0\to D^+D^-$:}
    The decay $B_d^0\to D^+D^-$ is a CP-even final state ($\eta_{\text{CP}}=+1$).
    We combine the 2022 average from HFLAV \cite{HFLAV:2022pwe}
\begin{align}
    \mathcal{A}_{\text{dir}}^{\text{CP}}(B_d^0\to D^+D^-) & = -0.13 \pm 0.12\:, \\
    \eta_{\text{CP}}\mathcal{A}_{\text{mix}}^{\text{CP}}(B_d^0\to D^+D^-) & = \phantom{-}0.84 \pm 0.10\:,
\end{align}
    taking into account $\rho=-0.18$, with the new measurement from LHCb \cite{LHCb:2024gkk}
\begin{align}
    \mathcal{A}_{\text{dir}}^{\text{CP}}(B_d^0\to D^+D^-) & = 0.128 \pm 0.103 \text{(stat)} \pm 0.010 \text{(syst)}\:, \\
    \eta_{\text{CP}}\mathcal{A}_{\text{mix}}^{\text{CP}}(B_d^0\to D^+D^-) & = 0.552 \pm 0.100 \text{(stat)} \pm 0.010 \text{(syst)}\:,
\end{align}
    taking into account $\rho=-0.472$.
    The new LHCb measurement uses the ``Run 2" data sample corresponding to an integrated luminosity of $6\text{fb}^{-1}$.
    \item \textbf{Direct CP asymmetry of the decay $B^+\to J/\psi\pi^+$:}
    The direct CP asymmetry in the decay $B^+\to J/\psi\pi^+$ has been measured by the BaBar, Belle, D0 and LHCb experiments.
    The experimental average is fully dominated by the recently released measurement from the LHCb collaboration \cite{LHCb:2024exp}
\begin{align}
    \mathcal{A}_{\text{dir}}^{\text{CP}}(B^+\to J/\psi\pi^+) & = \left(1.51 \pm 0.50\text{(stat)} \pm 0.08 \text{(syst)}\right)\times 10^{-2}\:.
\end{align}
    This result is obtained using a data sample corresponding to an integrated luminosity of 8.4 fb$^{-1}$.
    \item \textbf{Direct CP asymmetry of the decay $B^+\to J/\psi K^+$:}
    The direct CP asymmetry in the decay $B^+\to J/\psi K^+$ has been measured by the CLEO2, BaBar, Belle, D0 and LHCb experiments.
    The average provided by the PDG \cite{PDG:2024cfk} is
\begin{align}
    \mathcal{A}_{\text{dir}}^{\text{CP}}(B^+\to J/\psi K^+) & = \left(1.8 \pm 3.0\right)\times 10^{-3}\:.
\end{align}
    \item \textbf{Direct CP asymmetry of the decay $B_s^0\to J/\psi\bar K^{*0}$:}
    The polarisation-dependent direct CP asymmetries of the decay $B_s^0\to J/\psi\bar K^{*0}$ have been measured by the LHCb experiment \cite{LHCb:2015esn} using the ``Run 1" data sample, corresponding to an integrated luminosity of 3 fb$^{-1}$.
    To include in our fit framework, we weight the polarisation-dependent results according to the measured polarisation fractions to obtain a polarisation-independent result
\begin{align}
    \mathcal{A}_{\text{dir}}^{\text{CP}}(B_s^0\to J/\psi\bar K^{*0}) & = \left(0.9 \pm 5.2\right)\times 10^{-2}\:.
\end{align}
\end{itemize}

These inputs are also summarised in Table \ref{tab:inputs_future}.

The ``Pseudoscalar", ``Vector" and ``$B\to DD$" fit strategies only determine one of the two $B_q^0$--$\bar B_q^0$ mixing phases, while the mixing-induced CP asymmetry observables of at least one of the control modes in the fit depends on the other mixing phase.
These strategies thus also require external input for either $\phi_d$ or $\phi_s$.
In these three specific fit strategies we assume for simplicity that $\phi_q^{\text{eff}}=\phi_q$, ignoring the impact of the hadronic phase shift $\Delta\phi_q$.
We can then directly use the experimental measurements \eqref{eq:phid_eff} and \eqref{eq:phis_eff} for the effective mixing phases as the needed external constraints for $\phi_d$ and $\phi_s$, respectively.
The ``$B\to J/\psi X$", ``Nominal" and ``Extended" fit strategies determine $\phi_d$ or $\phi_s$ simultaneously and thus do not require these external inputs.

\section{Results from Current Data}\label{sec:current}
For each of the six different fit strategies, we use the GammaCombo framework \cite{LHCb:2016mag} to simultaneously fit for the $B_q^0$--$\bar B_q^0$ mixing phases $\phi_d$ and $\phi_s$ as well as the relevant penguin parameters.
Each fit strategy take as inputs the direct and mixing-induced CP asymmetries listed above, combined with external constraints on the CKM angle $\gamma$ and the Wolfenstein parameter $\lambda$.

\subsection[B2JpsiX Fit]{$B\to J/\psi X$ Fit}\label{sec:B2JpsiX}
\begin{figure}
    \centering
    \includegraphics[width=0.49\textwidth]{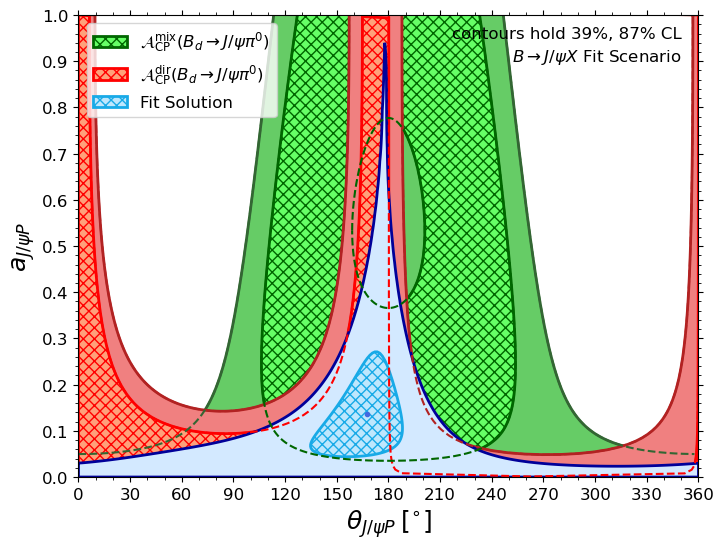}
    \includegraphics[width=0.49\textwidth]{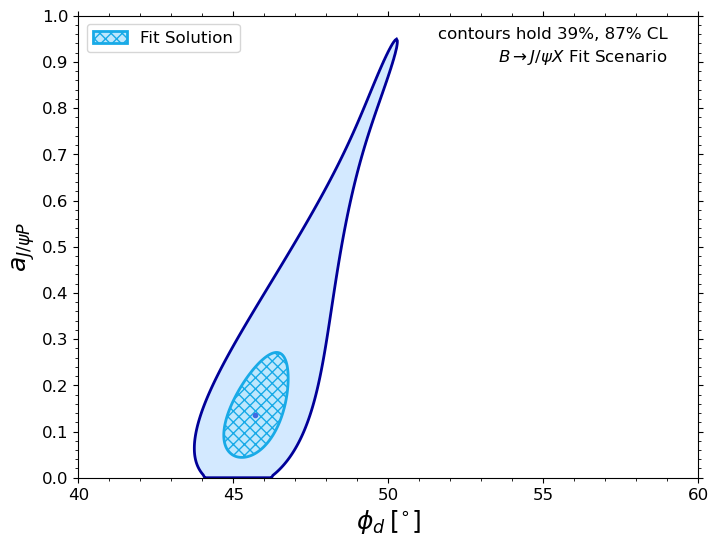}
    
    \includegraphics[width=0.49\textwidth]{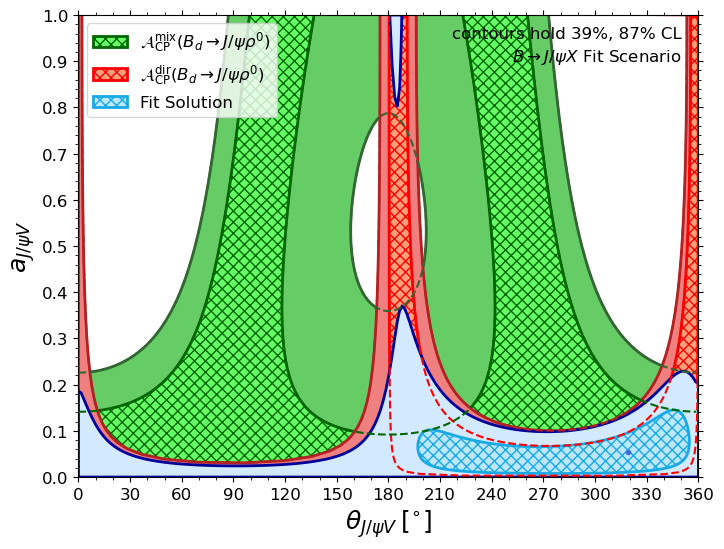}
    \includegraphics[width=0.49\textwidth]{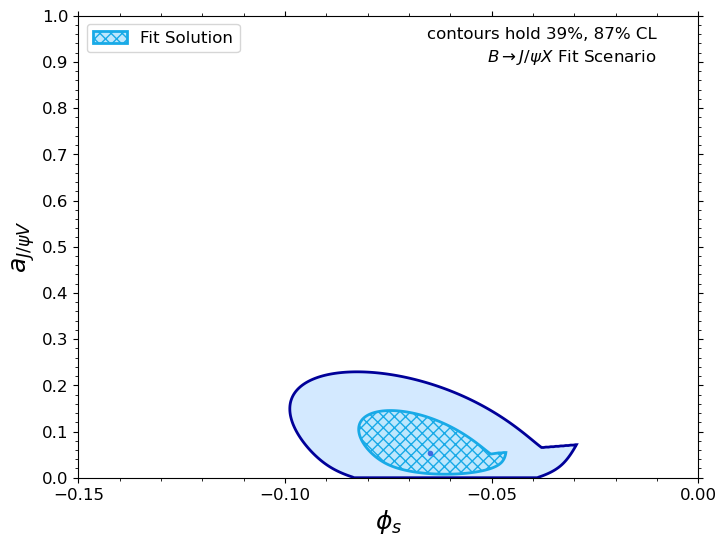}
    \caption{Two-dimensional confidence regions of the fit for the penguin parameters, $\phi_d$ and $\phi_s$ from the CP asymmetries in the $B\to J/\psi X$ decays.
    Note that the contours for $\mathcal{A}_{\text{CP}}^{\text{dir}}$ and $\mathcal{A}_{\text{CP}}^{\text{mix}}$ are added for illustration only.
    They include the best fit solutions for $\phi_d$, $\phi_s$ and $\gamma$ as Gaussian constraints.
    }
    \label{fig:Autumn24}
\end{figure}

\begin{figure}
    \centering
    \includegraphics[width=0.49\textwidth]{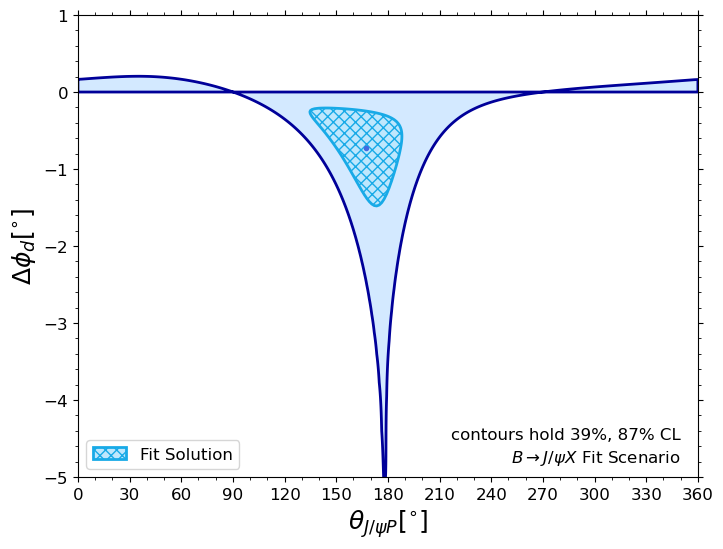}
    \includegraphics[width=0.49\textwidth]{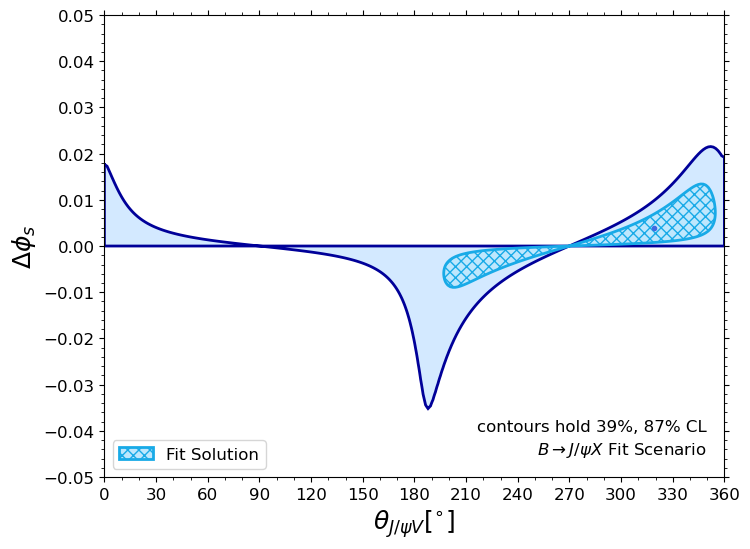}
    \caption{Two-dimensional confidence regions for the penguin shift $\Delta\phi$ as a function of penguin parameter $\theta$, corresponding to the fit to the CP asymmetries in the $B\to J/\psi X$ decays.
    }
    \label{fig:Autumn24_deltaphi}
\end{figure}

From the $B\to J/\psi X$ fit, we obtain the following values for the $B_q^0$--$\bar B_q^0$ mixing phases:
\begin{equation}
    \phi_d = \left(45.7_{-1.0}^{+1.1}\right)^{\circ}
    \qquad
    \phi_s = -0.065_{-0.017}^{+0.018} = \left(-3.72_{-0.97}^{+1.03}\right)^{\circ}\:,
\end{equation}
which take into account the effects due to penguin topologies.
The results for the penguin parameters are:
\begin{align}
    a_{J/\psi P} & = 0.14_{-0.09}^{+0.14} &
    \theta_{J/\psi P} & = \left(167_{-32}^{+21}\right)^{\circ} \\
    a_{J/\psi V} & = 0.054_{-0.047}^{+0.091} &
    \theta_{J/\psi V} & = \left(319_{-122}^{+41}\right)^{\circ}\:.
\end{align}
The two-dimensional confidence level contours for $(\theta_{J/\psi P}, a_{J/\psi P})$, $(\theta_{J/\psi V}, a_{J/\psi V})$, $(\phi_d, a_{J/\psi P})$ and $(\phi_s, a_{J/\psi V})$ are shown in Fig.\ \ref{fig:Autumn24}.
The confidence level contours for the penguin parameters can be converted into confidence level contours for the penguin shifts $\Delta\phi_d$ and $\Delta\phi_s$, which are shown in Fig.\ \ref{fig:Autumn24_deltaphi}.


The bottom left panel in Fig.\ \ref{fig:Autumn24} shows that the CP asymmetries in $B_d^0\to J/\psi\rho^0$ lead to two possible solutions for the penguin parameters $a_{J/\psi V}$ and $\theta_{J/\psi V}$.
The solution with $a_{J/\psi V}<0.5$ is the preferred minimum, but mathematically a second minimum exists for $a_{J/\psi V}>1$.
The latter solution corresponds to the scenario in which the penguin topologies have a larger contribution to the decay amplitude than the tree topology, which is strongly disfavoured theoretically.
Moreover, for this solution $\phi_s$ is larger than zero and therefore not visible in the bottom right panel.
For the penguin parameters $a_{J/\psi P}$ and $\theta_{J/\psi P}$ this second, disfavoured solution is excluded by the constraint coming from the mixing-induced CP asymmetry in $B_s^0\to J/\psi K_{\text{S}}^0$.

The top right panel in Fig.\ \ref{fig:Autumn24} clearly shows that the solution for $\phi_d$ is strongly correlated with the value of the penguin parameter $a_{J/\psi P}$, and it is thus of the utmost importance to pin down $a_{J/\psi P}$ as precisely as possible.
The large uncertainty on $a_{J/\psi P}$ obtained from our fit can be explained as follows.
The current experimental results on the direct CP asymmetry measurements in the control modes are compatible with zero, which results in degenerate solutions for $a_{J/\psi P}$ around $\theta_{J/\psi P}\approx 0^{\circ}$ and $\theta_{J/\psi P}\approx 180^{\circ}$.
Simultaneously, also the current mixing-induced CP asymmetry measurement in $B_d^0\to J/\psi\pi^0$ prefers a solution around $\theta_{J/\psi P}\approx 180^{\circ}$.

The bottom right panel in Fig.\ \ref{fig:Autumn24} shows a much weaker correlation between $\phi_s$ and the penguin parameter $a_{J/\psi V}$.
This is driven by the value of the mixing-induced CP asymmetry measurement in $B_d^0\to J/\psi\rho^0$.

\subsection[B2DD Fit]{$B\to DD$ Fit}\label{sec:B2DD}
\begin{figure}
    \centering
    \includegraphics[width=0.49\textwidth]{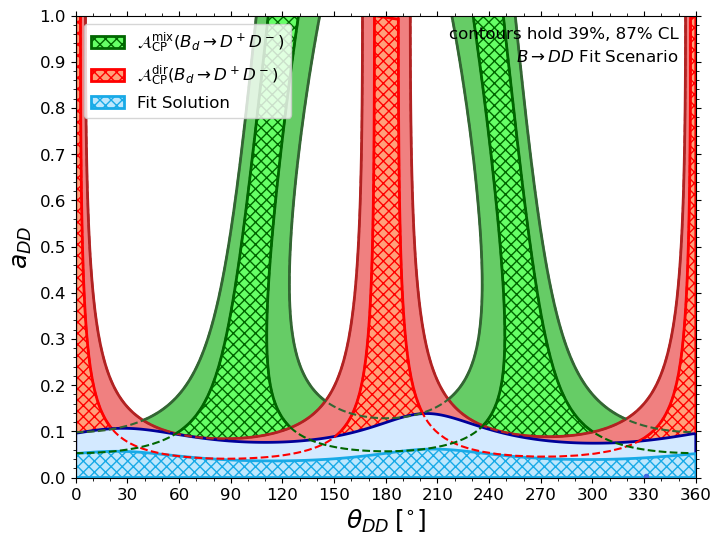}
    \includegraphics[width=0.49\textwidth]{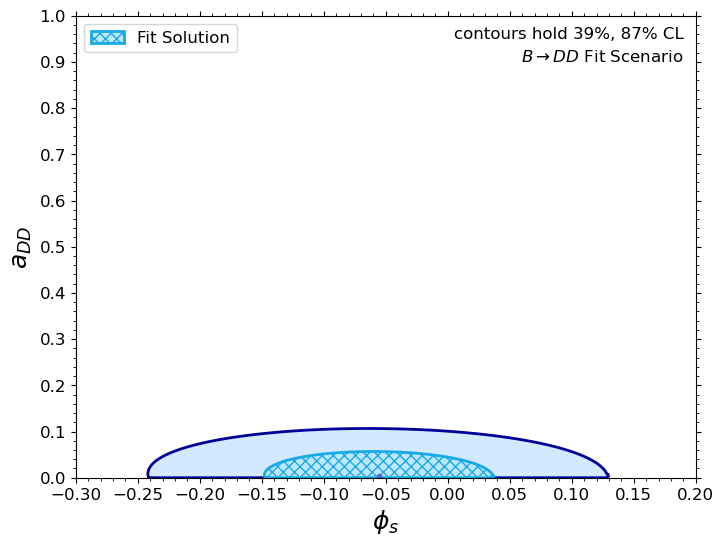}
    \caption{Two-dimensional confidence regions of the fit for the penguin parameters and $\phi_s$ from the CP asymmetries in the $B\to DD$ decays.
    Note that the contours for $\mathcal{A}_{\text{CP}}^{\text{dir}}$ and $\mathcal{A}_{\text{CP}}^{\text{mix}}$ are added for illustration only.
    They include the best fit solutions for $\phi_d$ and $\gamma$ as Gaussian constraints.
    }
    \label{fig:Autumn24_DD}
\end{figure}

The $B\to DD$ fit provides an alternative determination of the $B_s^0$--$\bar B_s^0$ mixing phase $\phi_s$
\begin{equation}
    \phi_s = -0.055_{-0.093}^{+0.092} = (-3.2 \pm 5.3)^{\circ}\:,
\end{equation}
which is fully consistent with the determination from the $B\to J/\psi X$ fit, but a factor 5 less precise.
The results for the penguin parameters are:
\begin{equation}
    a_{DD} = 0.003_{-0.003}^{+0.059}
    \qquad
    \theta_{DD} = \left(331_{-331}^{+29}\right)^{\circ}\:,
\end{equation}
where the phase $\theta_{DD}$ remains unconstrained because $a_{DD}$ is fully consistent with zero.
The two-dimensional confidence level contours for $(\theta_{DD}, a_{DD})$, $(\phi_s, a_{DD})$ are shown in Fig.~\ref{fig:Autumn24_DD}.

\begin{figure}
    \centering
    \includegraphics[width=0.49\textwidth]{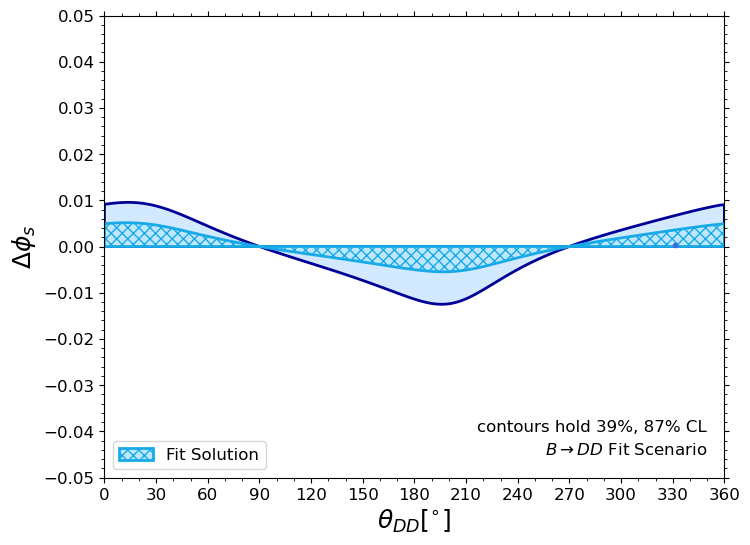}
    \caption{Two-dimensional confidence regions for the penguin shift $\Delta\phi_s$ as a function of penguin parameter $\theta$, corresponding to the fit to the CP asymmetries in the $B\to DD$ decays.
    }
    \label{fig:Autumn24_DD_deltaphi}
\end{figure}

\subsection{Nominal Fit}\label{sec:NomFit}
The inputs from the $B\to J/\psi X$ and $B\to DD$ fit strategies can be combined to perform one simultaneous fit for the $B_q^0$--$\bar B_q^0$ mixing phases $\phi_d$ and $\phi_s$ and the six penguin parameters.
This is our nominal fit strategy, and its results are:
\begin{equation}\label{eq:nominal_fit_I}
    \phi_d = \left(45.7_{-1.0}^{+1.1}\right)^{\circ}
    \qquad
    \phi_s = -0.065_{-0.017}^{+0.018} = (-3.72_{-0.97}^{+1.03})^{\circ}
\end{equation}
and
\begin{align}
    a_{J/\psi P} & = 0.14_{-0.09}^{+0.14} &
    \theta_{J/\psi P} & = \left(167_{-32}^{+21}\right)^{\circ} \\
    a_{J/\psi V} & = 0.053_{-0.046}^{+0.091} &
    \theta_{J/\psi V} & = \left(318_{-121}^{+42}\right)^{\circ} \\
    a_{DD} & = 0.008_{-0.008}^{+0.054} &
    \theta_{DD} & = \left(351_{-351}^{+9}\right)^{\circ}\:.\label{eq:nominal_fit_IV}
\end{align}
The fit results for $\phi_d$ and $\phi_s$ are identical to those from the $B\to J/\psi X$ fit.
The combination of the CP asymmetry measurements in the $B_s^0\to J/\psi\phi$ and $B_s^0\to D_s^+D_s^-$ decays does not lead to a better precision on $\phi_s$:
the precision is fully dominated by the mixing-induced CP asymmetry measurement in $B_s^0\to J/\psi\phi$.
The different central value for $\phi_s$ leads to a slightly larger value for the penguin parameter $a_{DD}$, while the other penguin parameters are identical to the results from the $B\to J/\psi X$ fit.

\begin{figure}
    \centering
    \includegraphics[width=0.49\textwidth]{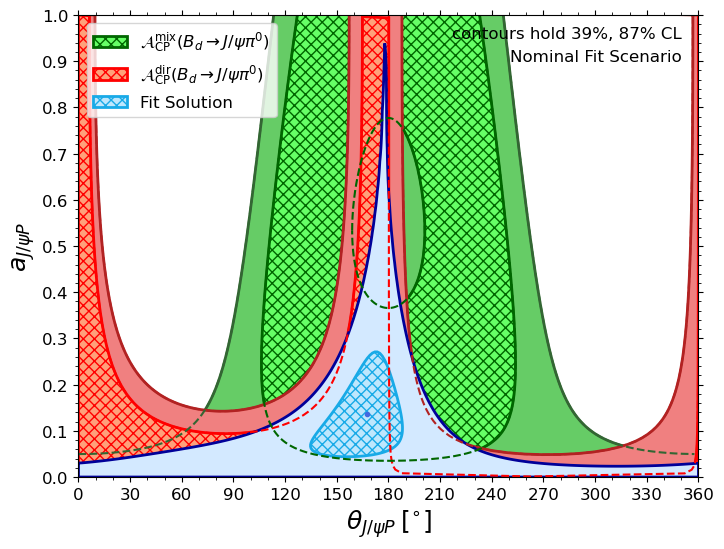}
    \includegraphics[width=0.49\textwidth]{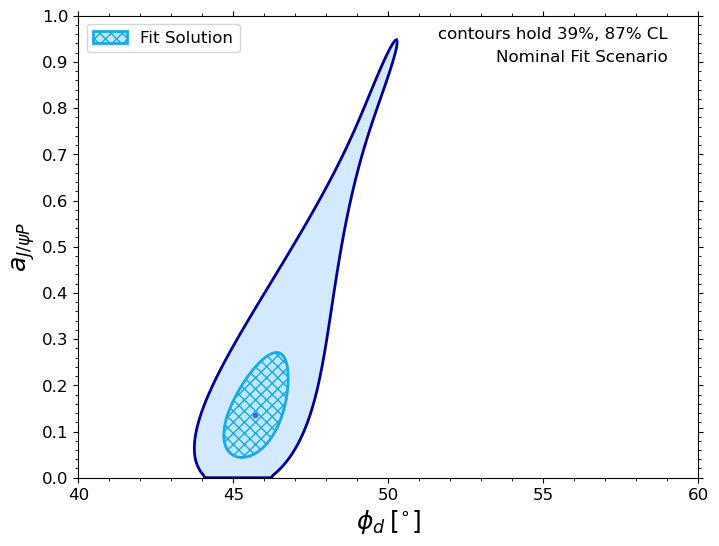}

    \includegraphics[width=0.49\textwidth]{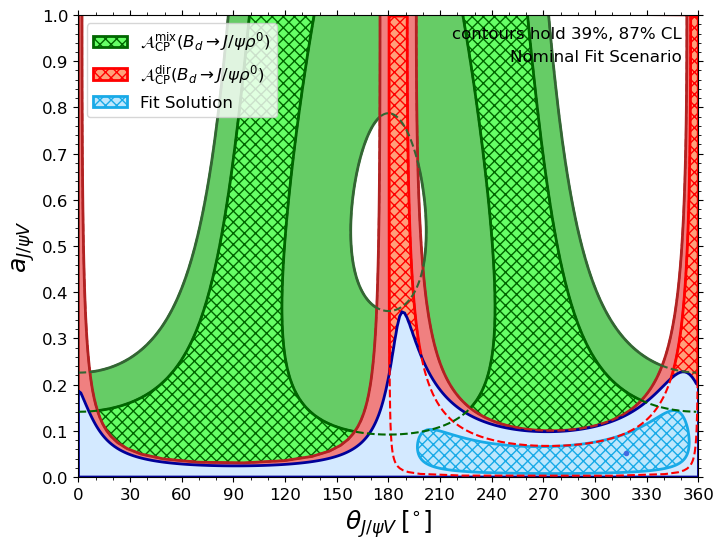}
    \includegraphics[width=0.49\textwidth]{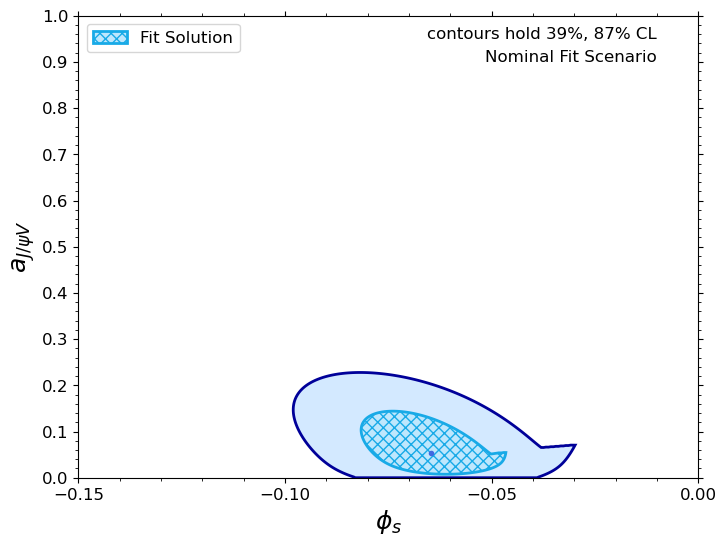}

    \includegraphics[width=0.49\textwidth]{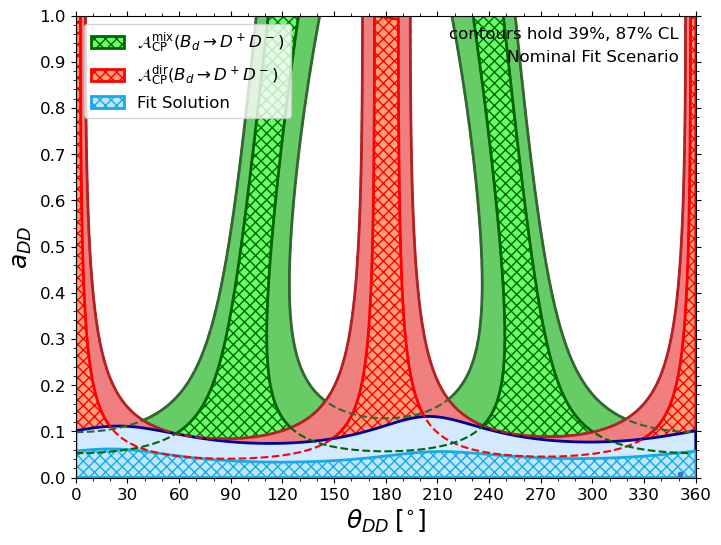}
    \includegraphics[width=0.49\textwidth]{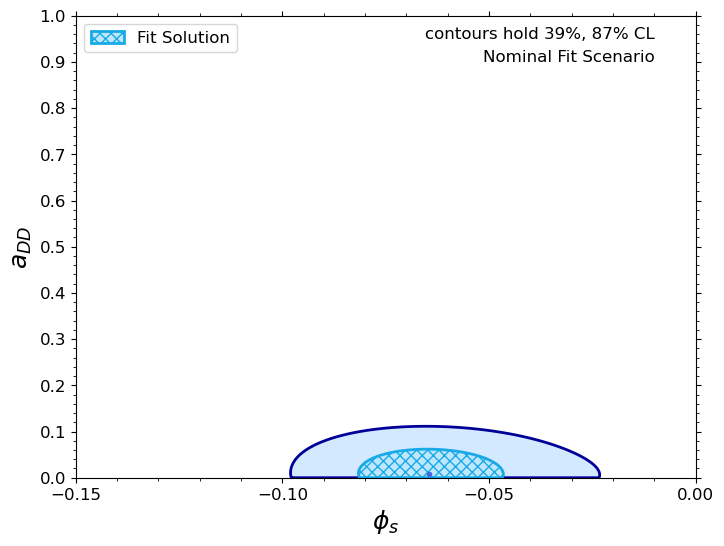}
    \caption{Two-dimensional confidence regions of our nominal fit for the penguin parameters, $\phi_d$ and $\phi_s$ from the CP asymmetries in the $B\to J/\psi X$ and $B\to DD$ decays.
    Note that the contours for $\mathcal{A}_{\text{CP}}^{\text{dir}}$ and $\mathcal{A}_{\text{CP}}^{\text{mix}}$ are added for illustration only.
    They include the best fit solutions for $\phi_d$, $\phi_s$ and $\gamma$ as Gaussian constraints.
    }
    \label{fig:Autumn24_nom}
\end{figure}

\begin{figure}
    \centering
    \includegraphics[width=0.49\textwidth]{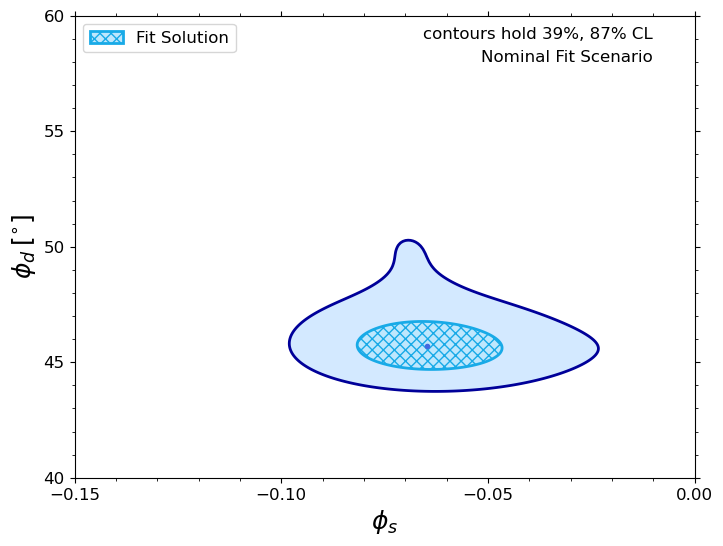}
    \caption{Two-dimensional confidence regions of our nominal fit for the penguin parameters, $\phi_d$ and $\phi_s$ from the CP asymmetries in the $B\to J/\psi X$ and $B\to DD$ decays.
    }
    \label{fig:Autumn24_nom2}
\end{figure}

The two-dimensional confidence level contours are shown in Fig.\ \ref{fig:Autumn24_nom}.
They are very similar to those from the $B\to J/\psi X$ fit in Fig. \ref{fig:Autumn24}, although the second, disfavoured solution for $(\theta_{J/\psi V},a_{J/\psi V})$ has now disappeared due to small numerical changes following the inclusion of the $B\to DD$ decays.
The correlation between the fit solutions for $\phi_d$ and $\phi_s$ is shown in Fig.\ \ref{fig:Autumn24_nom2}.
The latter shows that the fit results for $\phi_d$ and $\phi_s$ are uncorrelated.
The small bulge in the 87\% confidence level extending to $\phi_d>50^{\circ}$ is caused by the strong correlation of $\phi_d$ with the penguin parameter $a_{J/\psi P}$, shown in the top right panels of Figs.\ \ref{fig:Autumn24} and \ref{fig:Autumn24_nom}.


\subsection{Extended Fit}
\begin{figure}
    \centering
    \includegraphics[width=0.49\textwidth]{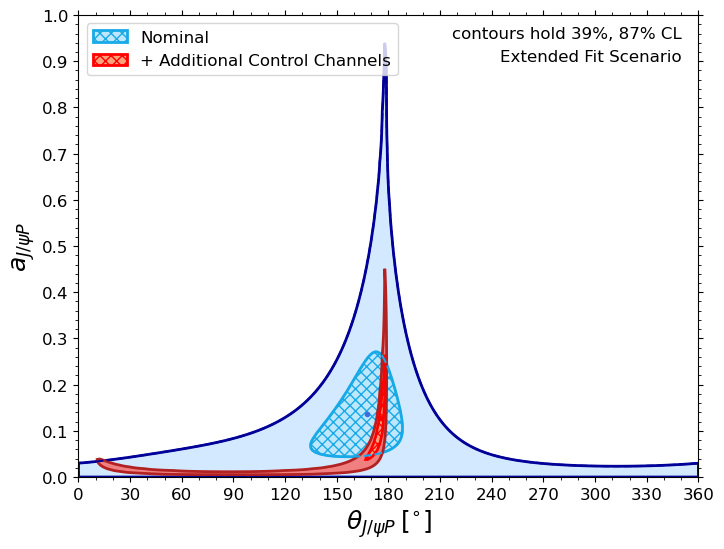}
    \includegraphics[width=0.49\textwidth]{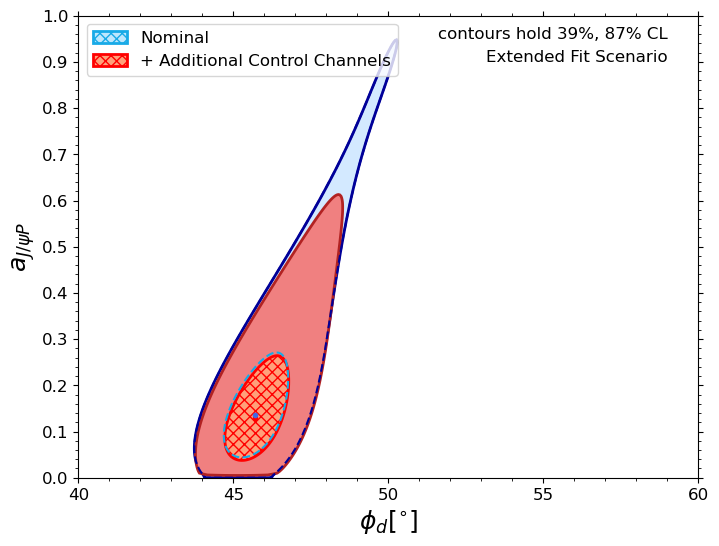}
    
    \includegraphics[width=0.49\textwidth]{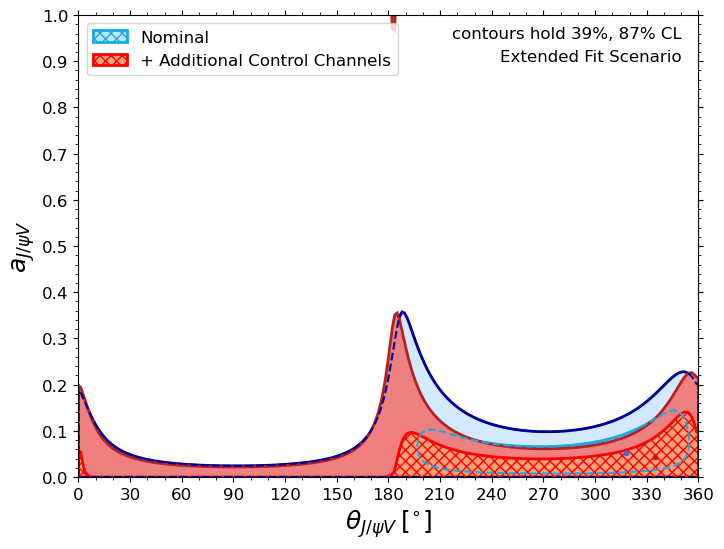}
    \includegraphics[width=0.49\textwidth]{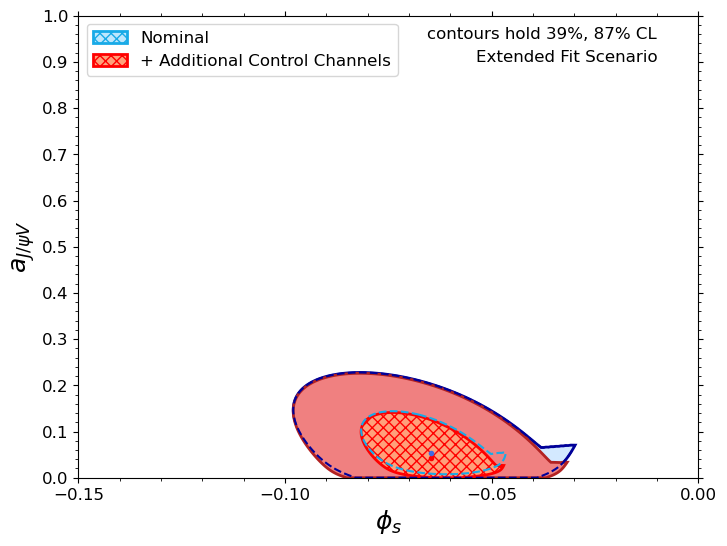}

    \includegraphics[width=0.49\textwidth]{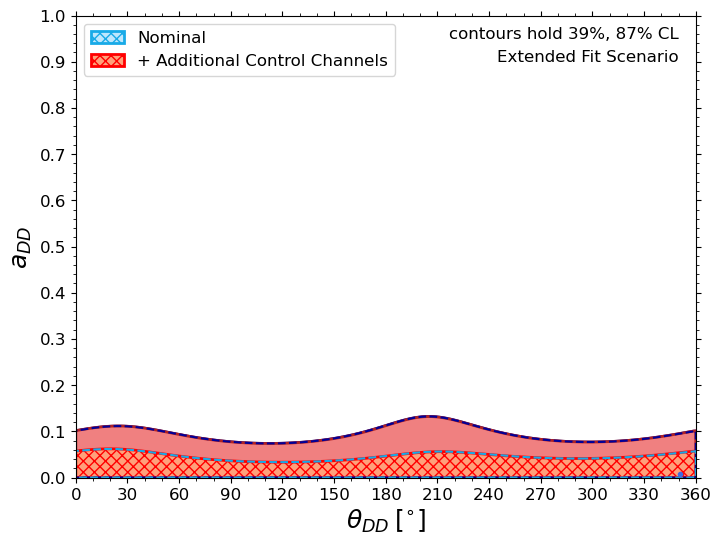}
    \includegraphics[width=0.49\textwidth]{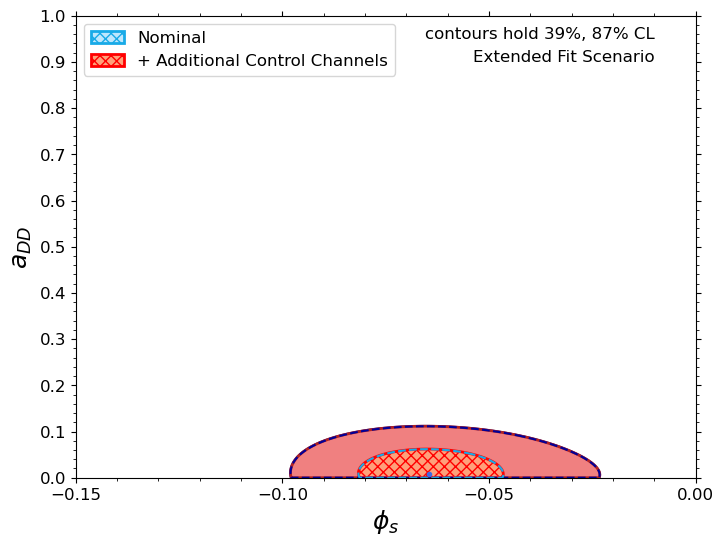}
    \caption{Two-dimensional confidence regions of the extended fit for the penguin parameters from the CP asymmetries in the $B\to J/\psi X$ and $B\to DD$ decays.
    }
    \label{fig:Autumn24_ext}
\end{figure}

The nominal fit can be extended with additional inputs from the direct CP asymmetries in $B^+\to J/\psi\pi^+$, $B^+\to J/\psi K^+$ and $B_s^0\to J/\psi\bar K^{*0}$.
The results of this extended fit are:
\begin{equation}
    \phi_d = \left(45.7_{-1.0}^{+1.1}\right)^{\circ}
    \qquad
    \phi_s = -0.065_{-0.017}^{+0.018} = (-3.72_{-0.97}^{+1.03})^{\circ}
\end{equation}
and
\begin{align}
    a_{J/\psi P} & = 0.13_{-0.09}^{+0.13} &
    \theta_{J/\psi P} & = \left(176.0_{-10.0}^{+0.6}\right)^{\circ} \\
    a_{J/\psi V} & = 0.044_{-0.044}^{+0.097} &
    \theta_{J/\psi V} & = \left(334_{-334}^{+26}\right)^{\circ} \\
    a_{DD} & = 0.008_{-0.008}^{+0.054} &
    \theta_{DD} & = \left(351_{-351}^{+9}\right)^{\circ}\:,
\end{align}
where the phases $\theta_{J/\psi V}$ and $\theta_{DD}$ are unconstrained because both $a_{J/\psi V}$ and $a_{DD}$ are fully consistent with zero in this fit strategy.
The two-dimensional confidence level contours shown in Fig.\ \ref{fig:Autumn24_ext}.

The inclusion of the three additional decay channels improves the precision on $\theta_{J/\psi P}$, but has no impact on the numerical fit results for $\phi_d$ and $\phi_s$.
The small value of the measured direct CP asymmetry in $B^+\to J/\psi\pi^+$ implies either $a_{J/\psi P}\approx 0$ or $\sin\theta_{J/\psi P}\approx 0$, but only the latter option is consistent with the measured value for the mixing-induced CP asymmetry in $B_d^0\to J/\psi\pi^0$.
This thus leads to strong constraints on $\theta_{J/\psi P}$, but not on $a_{J/\psi P}$ or $\phi_d$.

\subsection{Comparison}
The numerical results from all six fit strategies can be compared in Tab.\ \ref{tab:current_fits_ver}.

\begin{table}
    \resizebox{\textwidth}{!}
    {
    \renewcommand{\baselinestretch}{1.25}
    \selectfont
    \begin{tabular}{|c|c|c|c|c|c|c|}
        \cmidrule{1-4}
        Fit & $\phi_s$ & $\phi_s \:[^{\circ}]$ & $\phi_d \:[^{\circ}]$ \\
        \cmidrule{1-4}
Pseudoscalar &
\textcolor{red}{$-0.061 \pm 0.014$} & \textcolor{red}{$-3.50 \pm 0.80$} & $45.7_{-1.0}^{+1.1}$ \\
Vector &
$-0.064 \pm 0.018$ & $-3.67 \pm 1.03$ & \textcolor{red}{$45.12 \pm 0.94$} \\
$B\to J/\psi X$ &
$-0.065_{-0.017}^{+0.018}$ & $-3.72_{-0.97}^{+1.03}$ & $45.7_{-1.0}^{+1.1}$ \\
$B\to DD$ &
$-0.055_{-0.093}^{+0.092}$ & $-3.2 \pm 5.3$ & \textcolor{red}{$45.12 \pm 0.94$} \\
Nominal &
$-0.065_{-0.017}^{+0.018}$ & $-3.72_{-0.97}^{+1.03}$ & $45.7_{-1.0}^{+1.1}$ \\
Extended &
$-0.065_{-0.017}^{+0.018}$ & $-3.72_{-0.97}^{+1.03}$ & $45.7_{-1.0}^{+1.1}$ \\
        \toprule
        Fit & $a_{J/\psi P}$ & $\theta_{J/\psi P} \:[^{\circ}]$ & $a_{J/\psi V}$ & $\theta_{J/\psi V} \:[^{\circ}]$ & $a_{DD}$ & $\theta_{DD} \:[^{\circ}]$ \\
        \midrule
Pseudoscalar &
$0.14_{-0.09}^{+0.14}$ & $167_{-32}^{+21}$ & -- & -- & -- & -- \\
Vector &
-- & -- & $0.050_{-0.043}^{+0.090}$ & $314_{-118}^{+40}$ & -- & -- \\
$B\to J/\psi X$ &
$0.14_{-0.09}^{+0.14}$ & $167_{-32}^{+21}$ & $0.054_{-0.047}^{+0.091}$ & $319_{-122}^{+41}$ & -- & -- \\
$B\to DD$ &
-- & -- & -- & -- & $0.003_{-0.003}^{+0.059}$ & $331_{-331}^{+29}$ \\
Nominal &
$0.14_{-0.09}^{+0.14}$ & $167_{-32}^{+21}$ & $0.053_{-0.046}^{+0.091}$ & $318_{-121}^{+42}$ & $0.008_{-0.008}^{+0.054}$ & $351_{-351}^{+9}$ \\
Extended &
$0.13_{-0.09}^{+0.13}$ & $176.0_{-10.0}^{+0.6}$ & $0.044_{-0.044}^{+0.097}$ & $334_{-334}^{+26}$ & $0.008_{-0.008}^{+0.054}$ & $351_{-351}^{+9}$ \\
        \bottomrule
    \end{tabular}
    }
    \caption{Overview of the fit results from six different fit strategies.
    Quantities in red are included as external constraints.
    The ``--" indicates the parameter was not included in the fit.}
    \label{tab:current_fits_ver}
\end{table}

\section[SU3-Symmetry Breaking]{$SU(3)$-Symmetry Breaking}\label{sec:SU3}

The main systematic uncertainty associated with our fit framework is related to the violation of the $SU(3)$ flavour symmetry.
Generically, this symmetry is broken at the 20\% level, as illustrated by the ratios between the kaon and pion decay constants or between the $D_s$ and $D$ meson decay constants \cite{FLAG:2024oxs}.
However, $SU(3)$ symmetry breaking can be split between factorisable and non-factorisable contributions, where the latter are expected to be subdominant and no large effects have yet been observed \cite{Barel:2020jvf}.
Since we only include measurements of CP asymmetries in the various fit strategies, we only rely on the $SU(3)$ flavour symmetry assumption in Eq.\ \eqref{eq:SU3_pen}, which only get corrections from non-factorisable $SU(3)$-breaking effects.
We thus do not have to rely on Eq.\ \eqref{eq:SU3_had}, and are therefore not affected by the larger factorisable $SU(3)$-breaking effects.

By contrast, the experimental inputs from branching fraction measurements do rely on Eq.\ \eqref{eq:SU3_had} and are affected by both factorisable and non-factorisable $SU(3)$-breaking effects.
The use of branching fraction information to constrain the penguin effects in $\phi_d$ and $\phi_s$ has been explored in previous analyses \cite{DeBruyn:2014oga,Bel:2015wha}.
However, given the increased experimental precision on the CP asymmetries and in order to minimise the systematic uncertainty associated with $SU(3)$-breaking effects, we do not consider these inputs here.

In the nominal fit, we assume Eq.\ \eqref{eq:SU3_pen} to hold exactly.
In this section, we generalise this key relation to
\begin{equation}
    a' = a\times\xi_{SU(3)}^f\:,\qquad
    \theta' = \theta + \delta_{SU(3)}^f
\end{equation}
in order to study the impact of potential non-factorisable $SU(3)$-breaking effects on the obtained values of $\phi_d$ and $\phi_s$.
Here, $\xi_{SU(3)}^f$ is a scale factor and $\delta_{SU(3)}^f$ a phase difference parametrising the non-factorisable $SU(3)$-breaking effects between control mode $f$ and its $SU(3)$ partner.
In the absence of such effects $\xi_{SU(3)}^f=1$ and $\delta_{SU(3)}^f=0$.
In total, we introduce eight additional parameters, four scale factors $\xi_{SU(3)}^f$ and four phase differences $\delta_{SU(3)}^f$, linking the four pairs: $B_d^0\to J/\psi\pi^0$ and $B_d^0\to J/\psi K^0$, $B_s^0\to J/\psi K_{\text{S}}^0$ and $B_d^0\to J/\psi K^0$, $B_d^0\to J/\psi\rho^0$ and $B_s^0\to J/\psi\phi$, $B_d^0\to D^+D^-$ and $B_s^0\to D_s^+D_s^-$.

\begin{figure}
    \centering
    \includegraphics[width=0.49\textwidth]{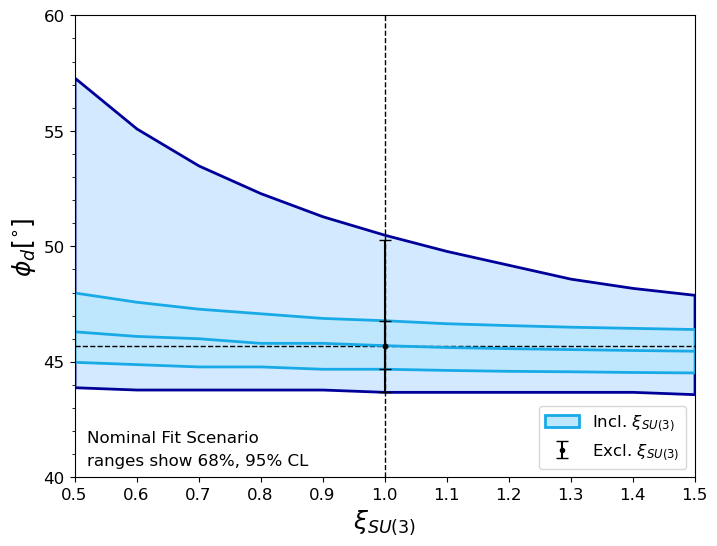}
    \includegraphics[width=0.49\textwidth]{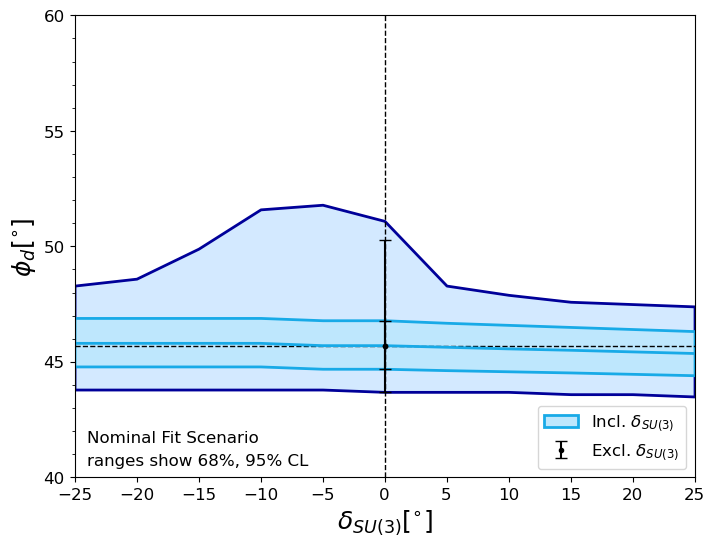}

    \includegraphics[width=0.49\textwidth]{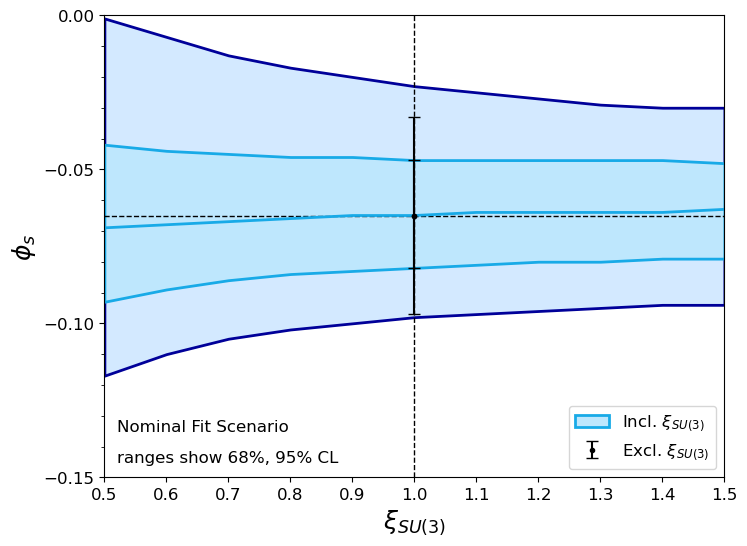}
    \includegraphics[width=0.49\textwidth]{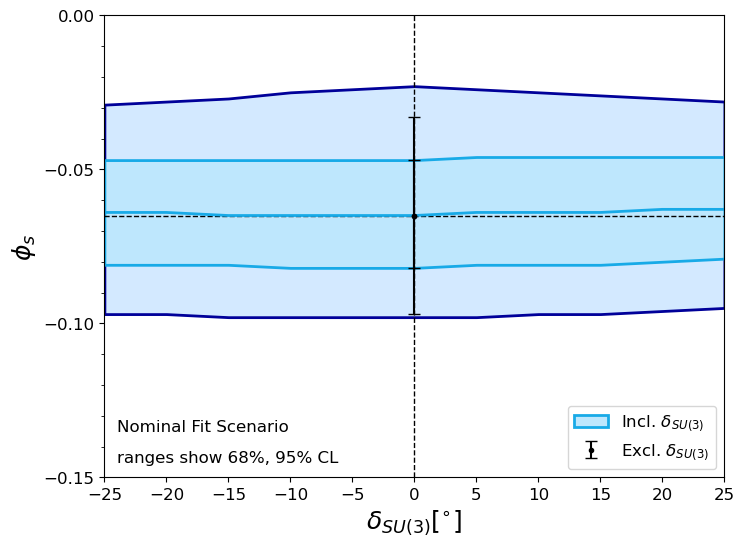}
    \caption{Dependence of the fit value and its uncertainties for $\phi_d$ (Top) and $\phi_s$ (Bottom) as function of the $SU(3)$-breaking parameters $\xi_{SU(3)}$ (Left) and $\delta_{SU(3)}$ (Right).
    The $SU(3)$-breaking parameters are identical for all four decay pairs and included as Gaussian constraints in the fit.
    The black data point shows the result from the nominal fit, in which exact $SU(3)$ symmetry is assumed.
    }
    \label{fig:phi_SU3_scan}
\end{figure}

\begin{figure}
    \centering
    \includegraphics[width=0.49\textwidth]{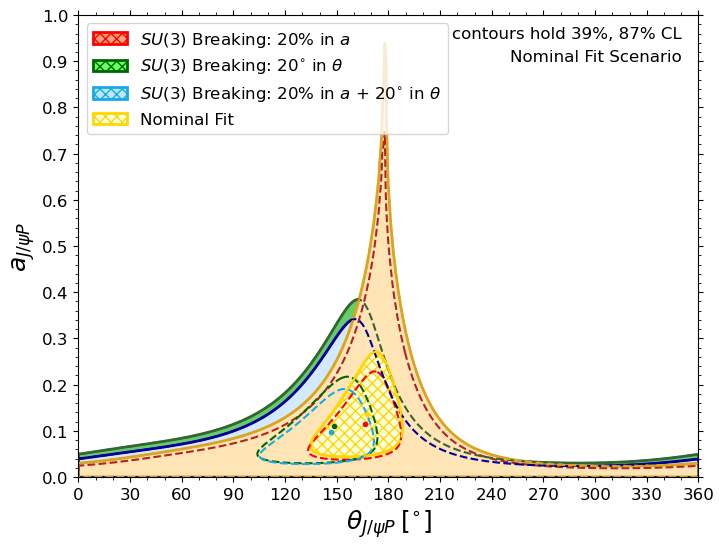}
    \includegraphics[width=0.49\textwidth]{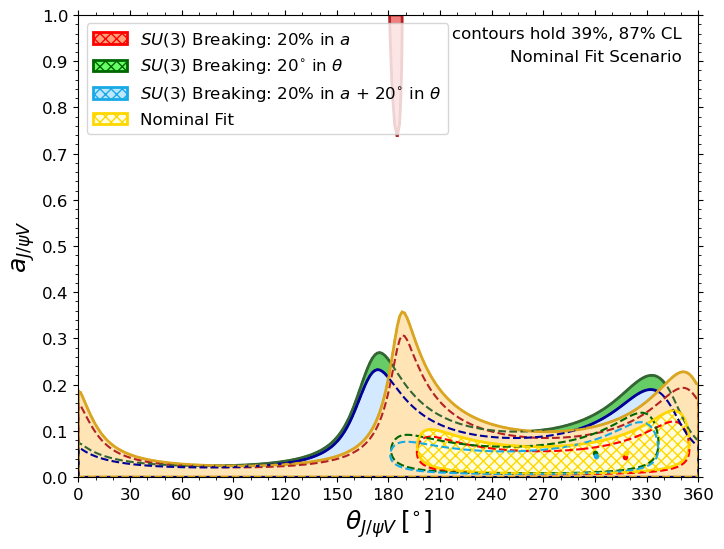}

    \includegraphics[width=0.49\textwidth]{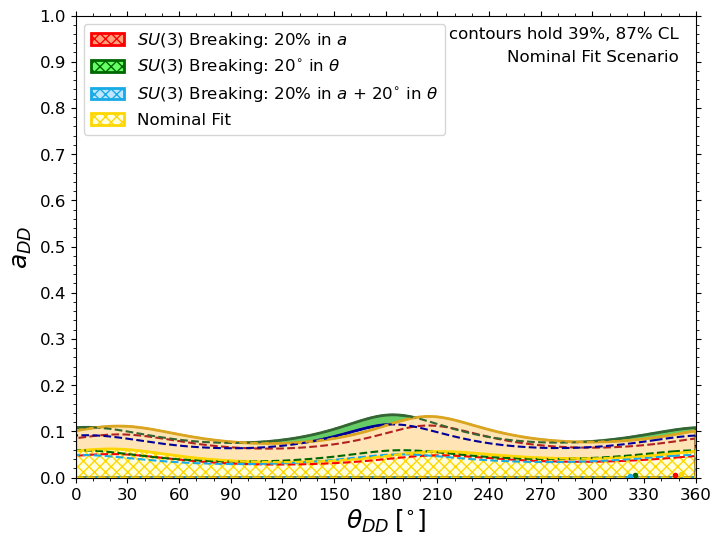}
    \includegraphics[width=0.49\textwidth]{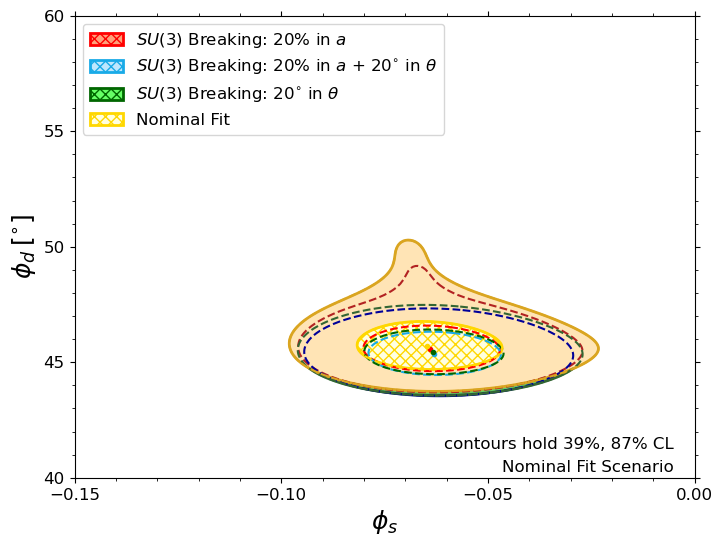}
    \caption{Two-dimensional confidence regions of the fit for $\phi_d$ and $\phi_s$ from the CP asymmetries in the $B\to J/\psi X$ and $B\to DD$ decays assuming different $SU(3)$-breaking scenarios.
    The $SU(3)$-breaking effects are included as Gaussian constraints in the fit.
    }
    \label{fig:phi_SU3}
\end{figure}

The fit framework, as presented in this paper, cannot simultaneously determine the penguin parameters $(a,\theta)$ and the $SU(3)$-breaking parameters $(\xi_{SU(3)},\delta_{SU(3)})$ from the experimental inputs as that leaves the fits underconstrained.
The $SU(3)$ breaking parameters therefore have to be provided as external inputs to the analyses.
The impact of potential non-factorisable $SU(3)$-breaking effects is explored by repeating the nominal fit for different values of $\xi_{SU(3)}^f$ and $\delta_{SU(3)}^f$.
For simplicity, we treat all four decay channel pairs equally and do not differentiate between the four scale factors or between the four phase differences, i.e.\ $\xi_{SU(3)}^f=\xi_{SU(3)}$ and $\delta_{SU(3)}^f=\delta_{SU(3)}$.
The impact on the fit values and uncertainties for $\phi_d$ and $\phi_s$ are shown in Fig.\ \ref{fig:phi_SU3_scan}.
For the plots on the left in Fig.\ \ref{fig:phi_SU3_scan}, we varied $\xi_{SU(3)}$ between 0.5 and 1.5 while keeping $\delta_{SU(3)}=0$ fixed.
The parameter $\xi_{SU(3)}$ is included as a Gaussian constraint with an uncertainty of 0.1.
For the plots on the right in Fig.\ \ref{fig:phi_SU3_scan}, we varied $\delta_{SU(3)}$ between $-25^{\circ}$ and $+25^{\circ}$ while keeping $\xi_{SU(3)}=1$ fixed.
The parameter $\delta_{SU(3)}$ is included as Gaussian constraint with an uncertainty of $5^{\circ}$.
For the specific scan points
\begin{equation}
    \xi_{SU(3)} = 1.2 \pm 0.1\:,
    \qquad\text{and/or}\qquad
    \delta_{SU(3)} = (20 \pm 5)^{\circ}
\end{equation}
we also illustrate the impact of potential non-factorisable $SU(3)$-breaking effects on the two-dimensional confidence regions for the penguin parameters, $\phi_d$ and $\phi_s$ in Fig.\ \ref{fig:phi_SU3}.

The results in Figs.\ \ref{fig:phi_SU3_scan} and \ref{fig:phi_SU3} show that the impact of $SU(3)$-breaking effects on the central value and the $1\sigma$ uncertainties (68\% confidence level interval in 1D, 39\% confidence level contour in 2D) are small.
For the current experimental precision, the $SU(3)$-breaking effects only become important for the $2\sigma$ uncertainties (95\% confidence level interval in 1D, 87\% confidence level contour in 2D).
These results thus illustrate that the determination of $\phi_d$ and $\phi_s$ from our fit framework is robust against $SU(3)$-breaking effects, even though their exact size remains unknown.


\section{Future Scenarios}\label{sec:future}
Based on the results from the nominal fit scenario one concludes that the penguin parameters $a_{J/\psi V}$ and $a_{DD}$ turn out to be favourably small.
This implies that also the hadronic shift $\Delta\phi_s$ is small, as illustrated in Figs.\ \ref{fig:Autumn24_deltaphi} and \ref{fig:Autumn24_DD_deltaphi}, and the impact of higher order corrections to the determination of $\phi_s$ is minimal.
However, these observations do not diminish the importance of controlling the penguin effects.
From Fig.\ \ref{fig:Autumn24_deltaphi} one also concludes that the uncertainty on $\Delta\phi_s$ is in the order of 0.01, and thus of similar size as the experimental uncertainty on $\phi_s^{\text{eff}}$.
This increases the uncertainty on $\phi_s$ by approximately a factor $\sqrt{2}$.
So even though the penguin effects are small, the uncertainty on $a_{J/\psi V}$ will soon become the dominant uncertainty on the determination of $\phi_s$ unless the measurements of the CP asymmetries in the control modes are further improved.
The penguin control modes should thus play a prominent role in the HL-LHC and Belle-II programmes in order to maximally benefit from the foreseen increased precision on the golden decay channels $B_d^0\to J/\psi K^0$ and $B_s^0\to J/\psi\phi$.
Let us therefore explore what future precision we can expect on $\phi_d$ and $\phi_s$ given different scenarios for the penguin control modes.

\paragraph{Fit Scenarios}
To illustrate how the current picture could improve with updated experimental inputs, we consider two time frames:
\begin{itemize}
    \item \textbf{End of Run 3:} At the end of LHC Run 3 in summer 2026, the LHCb experiment expects to have collected a dataset corresponding to an integrated luminosity of 23 fb$^{-1}$ \cite{LHCb:2018roe}.
    Simultaneously, the Belle-II experiment predicts to have collected 1 ab$^{-1}$ of data \cite{superkekb}.
    \item \textbf{End of HL-LHC:} At the end of the High-Luminosity LHC programme, the  LHCb experiment expects to have collected a dataset corresponding to an integrated luminosity of 300 fb$^{-1}$ \cite{LHCb:2018roe}.
    Simultaneously, the Belle-II experiment expects to complete its data taking campaign, having collected 50 ab$^{-1}$ of data \cite{Belle-II:2018jsg}.
    We assume the detector upgrade necessary to achieve this luminosity \cite{ATLAS:2025lrr} will happen.
\end{itemize}
To illustrate the importance of controlling the penguin contributions, we consider two scenarios:
\begin{itemize}
    \item \textbf{Excl.\ Penguins:} In this scenario we only consider updated measurements for the modes $B_d^0\to J/\psi K^0$, $B_s^0\to J/\psi\phi$ and $B_s^0\to D_s^+D_s^-$.
    We assume no new CP asymmetry measurements will become available for the penguin control modes $B_d^0\to J/\psi\pi^0$, $B_s^0\to J/\psi K_{\text{S}}^0$, $B_s^0\to J/\psi \rho^0$ and $B_d^0\to D^+D^-$.
    \item \textbf{Incl.\ Penguins:} In this scenario we assume the seven decay channels analysed in the nominal fit scenario receive equal attention and new CP asymmetry measurements will become available for all of them.
\end{itemize}
Together these options result in four future scenarios.

For all scenarios we make the same assumptions, listed in Section \ref{sec:fits}, as used for the nominal fit to the current experimental data.
In particular, this means that $SU(3)$-symmetry breaking effects are not included.
But as demonstrated in the previous section, our fit framework is robust against these corrections, which have a minimal impact on the results for $\phi_d$ and $\phi_s$.

To simplify interpretation of the results, we recalculate the central values for all CP asymmetry observables to correspond to the solution obtained from the nominal fit, i.e. Eqs.\ \eqref{eq:nominal_fit_I}--\eqref{eq:nominal_fit_IV}.
This avoids artificial tensions between the input observables which could arise from reducing the uncertainties when the central values currently do not align due to statistical fluctuations.
Moreover, the best fit solution stays the same for all scenarios, which allows for an easy comparison of their results.
Note, however, that because of the shifted central values the confidence level contours representing the current precision in the plots shown below are slightly different from the results shown in Figs.\ \ref{fig:Autumn24_nom} and \ref{fig:Autumn24_nom2}.

For the expected uncertainties on the input observables we use the official prospects published by LHCb \cite{LHCb:2018roe} and Belle-II \cite{Belle-II:2018jsg} where possible.
Based on recent measurements published by LHCb and Belle-II, some of these projections are conservative, especially regarding systematic uncertainties.
In view of the European Strategy for Particle Physics Update, an updated overview of the prospects has also been compiled in Ref.\ \cite{ATLAS:2025lrr}.
When no official prospects are available, which is often the case for the control modes, we extrapolate the current statistical uncertainty with the expected increase in luminosity, and conservatively assume the current systematic uncertainty to be fully irreducible.
In reality the systematic uncertainty consists of both reducible and irreducible components, but disentangling them goes beyond the scope of this study.
Because we choose not to scale the systematic uncertainty, multiple input observables will become systematics limited in our future scenarios.
When scaling LHCb results that only use Run 1 data, we assume a factor 2 increase in the production cross-section $\sigma_{b\bar b}$ between Run 1 and the subsequent data taking periods to reflect the increase in centre-of-mass energy from 7+8 TeV to 13 TeV.

\paragraph{Experimental Inputs}
The fit scenarios use the following inputs:
\begin{itemize}
    \item \textbf{External constraint on the CKM element $|V_{us}| = \lambda$:}
    We do not consider improvements to this input observables.
    \item \textbf{External constraint on the CKM angle $\gamma$:}
    LHCb expects to reach a precision on $\gamma$ from $B\to DK$ decays of $1.5^{\circ}$ at the end of LHC Run 3, and $0.35^{\circ}$ at the end of the HL-LHC programme \cite{LHCb:2018roe}.
    Belle-II expects to reach a precision of $1.6^{\circ}$ with 50 ab$^{-1}$ \cite{Belle-II:2018jsg}.
    For the \emph{End of Run 3} scenario, we extrapolate this number back to 1 ab$^{-1}$, resulting in an expected precision of $11^{\circ}$.
    We use the weighted average of the LHCb and Belle-II prospects for our future scenarios.
    \item \textbf{CP asymmetries of the decay $B_d^0\to J/\psi K^0$:}
    LHCb expects to reach a precision on ``$\sin2\beta$" from $B_d^0\to J/\psi K^0$ of 0.011 at the end of LHC Run 3, and 0.003 at the end of the HL-LHC programme \cite{LHCb:2018roe}, which we interpret as estimates for the mixing-induced CP asymmetry measurement.
    No explicit estimates for the direct CP asymmetry are given, which we therefore scale based on the Run 2 measurement \cite{LHCb:2023zcp} given in Eq.\ \eqref{eq:Adir_Bd2JpsiK}.
    For Belle-II, we use the prospects from Ref.\ \cite{Belle-II:2018jsg}.
    Since the systematic uncertainty on the current measurement is already better than the ``No improvement" scenario given in Ref.\ \cite{Belle-II:2018jsg}, we use that instead.
    For the \emph{End of Run 3} scenario, we extrapolate the statistical uncertainty of the Belle-II prospects back to 1 ab$^{-1}$, while keeping the same systematic uncertainty.
    \item \textbf{CP asymmetries of the decay $B_s^0\to J/\psi K_{\text{S}}^0$:}
    No prospects are available from LHCb or Belle-II, so we scale the statistical uncertainty of the LHCb Run 1 measurement \cite{LHCb:2015brj} given in Eqs.\ \eqref{eq:Adir_Bs2JpsiKs} and \eqref{eq:Amix_Bs2JpsiKs}.
    \item \textbf{CP asymmetries of the decay $B_d^0\to J/\psi\pi^0$:}
    In Ref.\ \cite{Belle-II:2018jsg} Belle-II has published multiple prospects for a future measurement using the full 50 ab$^{-1}$ data set.
    Their expected precision on the CP asymmetries will be limited by the irreducible systematic uncertainty, which is estimated to be between 0.04 and 0.05.
    However, their current CP asymmetry measurements in Eqs.\ \eqref{eq:Adir_Bd2JpsiPi_B2} and \eqref{eq:Amix_Bd2JpsiPi_B2} already outperform these prospects.
    We therefore take the systematic uncertainty from Eqs.\ \eqref{eq:Adir_Bd2JpsiPi_B2} and \eqref{eq:Amix_Bd2JpsiPi_B2} and scale the statistical uncertainty with the expected increase in luminosity.
    \item \textbf{CP asymmetries of the decay $B_s^0\to J/\psi\phi$:}
    LHCb expects to reach a precision on $\phi_s^{\text{eff}}$ of 14 mrad at the end of LHC Run 3, and 4 mrad at the end of the HL-LHC programme \cite{LHCb:2018roe}.
    We recalculate the uncertainty of the weighted average using the published ATLAS, CDF, CMS and D0 measurements, and these prospects from LHCb.
    For the ratio $\lambda_{J/\psi\phi}$, we recalculate the uncertainty of the weighted average using the published CMS measurement and scaling the uncertainty of the LHCb measurement.
    These two inputs are used to recalculate the CP asymmetries.
    \item \textbf{CP asymmetries of the decay $B_d^0\to J/\psi \rho^0$:}
    No prospects are available from LHCb or Belle-II, so we scale the statistical uncertainty of the LHCb Run 1 measurement \cite{LHCb:2014xpr} given in Eqs.\ \eqref{eq:Adir_Bd2JpsiRho} and \eqref{eq:Amix_Bd2JpsiRho}.
    We again restrict our analysis to only use the polarisation-independent measurement.
    \item \textbf{CP asymmetries of the decay $B_s^0\to D_s^+D_s^-$:}
    LHCb expects to reach a precision on $\phi_s^{\text{eff}}$ of 35 mrad at the end of LHC Run 3, and 9 mrad at the end of the HL-LHC programme \cite{LHCb:2018roe}.
    For the ratio $\lambda_{D_sD_s}$, we scale the statistical uncertainty of the LHCb measurement.
    These two inputs are used to recalculate the CP asymmetries.
    \item \textbf{CP asymmetries of the decay $B_d^0\to D^+D^-$:}
    No prospects are available from LHCb or Belle-II, so we scale the statistical uncertainties of the current LHCb and Belle measurements, where the latter is used to estimate a future measurement from Belle-II.
    We recalculate the uncertainty of the weighted average using the published BaBar and Belle measurement, and these prospects for LHCb and Belle-II.
\end{itemize}
These inputs are also summarised in Table \ref{tab:inputs_future}.

\begin{sidewaystable}
    \footnotesize
    \centering
    \begin{tabular}{|c|c|c|c|c|c|}
        \toprule
        Observable & Current Value & Source & End of Run 3 & End of HL-LHC & Source \\
        \midrule
        $\lambda$ & 
        $0.22308 \pm 0.00055$ & \cite{Seng:2022wcw} & & & \\
        $\gamma$ & 
        $(65.6_{-3.0}^{+2.9})^{\circ}$ & HFLAV & 
        $(65.6\pm 1.5)^{\circ}$ & 
        $(65.6 \pm 0.34)^{\circ}$ & \cite{LHCb:2018roe,Belle-II:2018jsg} \\
        \midrule
    	$\phantom{\eta_{\text{CP}}}\mathcal{A}_{\text{dir}}^{\text{CP}}(B_d^0\to J/\psi K^0)$ & 
        $-0.007 \pm 0.012 \pm 0.014$ & \cite{HFLAV:2022pwe} & & & \\
    	$\eta_{\text{CP}}\mathcal{A}_{\text{mix}}^{\text{CP}}(B_d^0\to J/\psi K^0)$ & 
        $\phantom{-}0.690 \pm 0.017 \pm 0.006$ & \cite{HFLAV:2022pwe} & & & \\
        \midrule
    	$\phantom{\eta_{\text{CP}}}\mathcal{A}_{\text{dir}}^{\text{CP}}(B_d^0\to J/\psi K^0)$ & 
        $-0.035 \pm 0.026 \pm 0.029$ & \cite{Belle-II:2024lwr} & 
        $-0.003 \pm 0.018 \pm 0.029$ & $-0.0030 \pm 0.0025 \pm 0.0290$ & \cite{Belle-II:2018jsg}\\
    	$\eta_{\text{CP}}\mathcal{A}_{\text{mix}}^{\text{CP}}(B_d^0\to J/\psi K^0)$ & 
        $\phantom{-}0.724 \pm 0.035 \pm 0.009$ & \cite{Belle-II:2024lwr} & 
        $\phantom{-}0.707\phantom{0} \pm 0.025\phantom{0} \pm 0.008\phantom{0}$ & $\phantom{-}0.7065 \pm 0.0035 \pm 0.0083$ & \cite{Belle-II:2018jsg}\\
	    $\rho$ & $+0.09$ & \cite{Belle-II:2024lwr} & $+0.09$ & $+0.09$ & Copy\\
	    \midrule
    	$\phantom{\eta_{\text{CP}}}\mathcal{A}_{\text{dir}}^{\text{CP}}(B_d^0\to J/\psi K^0)$ & 
        $\phantom{-}0.015 \pm 0.013 \pm 0.003$ & \cite{LHCb:2023zcp} & 
        $-0.0030 \pm 0.008 \pm 0.003$ & 
        $-0.0030 \pm 0.002 \pm 0.003$ & Lumi\\
    	$\eta_{\text{CP}}\mathcal{A}_{\text{mix}}^{\text{CP}}(B_d^0\to J/\psi K^0)$ & 
        $\phantom{-}0.722 \pm 0.014 \pm 0.007$ & \cite{LHCb:2023zcp} & 
        $\phantom{-}0.707\phantom{0} \pm 0.011 \phantom{\pm 0.011}$ & 
        $\phantom{-}0.707\phantom{0} \pm 0.003 \phantom{\pm 0.011}$ & \cite{LHCb:2018roe} \\
    	$\rho$ & $+0.437$ & \cite{LHCb:2023zcp} & $+0.437$ & $+0.437$ & Copy\\
        \midrule
        $\phantom{\eta_{\text{CP}}}\mathcal{A}_{\text{dir}}^{\text{CP}}(B_s^0\to J/\psi K_{\text{S}}^0)$ & 
        $-0.28 \pm 0.41 \pm 0.08$ & \cite{LHCb:2015brj} & 
        $0.051 \pm 0.11 \pm 0.08$ & 
        $0.051 \pm 0.03 \pm 0.08$ & Lumi \\
        $\eta_{\text{CP}}\mathcal{A}_{\text{mix}}^{\text{CP}}(B_s^0\to J/\psi K_{\text{S}}^0)$ & 
        $\phantom{-}0.08 \pm 0.40 \pm 0.08$ & \cite{LHCb:2015brj} & 
        $0.17\phantom{0} \pm 0.10 \pm 0.08$ & 
        $0.17\phantom{0} \pm 0.03 \pm 0.08$ & Lumi\\
        $\rho$ & $+0.06$ & \cite{LHCb:2015brj} & $+0.06$ & $+0.06$ & Copy \\
        \midrule
        $\phantom{\eta_{\text{CP}}}\mathcal{A}_{\text{dir}}^{\text{CP}}(B_d^0\to J/\psi\pi^0)$ & 
        $0.04 \pm 0.12$ & \cite{HFLAV:2022pwe} & & & \\
        $\eta_{\text{CP}}\mathcal{A}_{\text{mix}}^{\text{CP}}(B_d^0\to J/\psi\pi^0)$ & 
        $0.86 \pm 0.14$ & \cite{HFLAV:2022pwe} & & & \\
        $\rho$ & $-0.08$ & \cite{HFLAV:2022pwe} & & & \\
        \midrule
        $\phantom{\eta_{\text{CP}}}\mathcal{A}_{\text{dir}}^{\text{CP}}(B_d^0\to J/\psi\pi^0)$ & 
        $0.13 \pm 0.12 \pm 0.03$ & \cite{Belle-II:2024hqw} & 
        $0.051 \pm 0.072 \pm 0.034$ & $0.051 \pm 0.010 \pm 0.034$ & Lumi\\
        $\eta_{\text{CP}}\mathcal{A}_{\text{mix}}^{\text{CP}}(B_d^0\to J/\psi\pi^0)$ & 
        $0.88 \pm 0.17 \pm 0.03$ & \cite{Belle-II:2024hqw} & 
        $0.86 \pm 0.10 \pm 0.03$ & $0.857 \pm 0.015 \pm 0.032$ & Lumi\\
        $\rho$ & $+0.08$ & \cite{Belle-II:2024hqw} & $+0.08$ & $+0.08$ & Copy\\
	    \midrule
    	$\phantom{\eta_{\text{CP}}}\mathcal{A}_{\text{dir}}^{\text{CP}}(B_s^0\to J/\psi \phi)$ & 
        $\phantom{-}0.005 \pm 0.009$ & This & 
        $\phantom{-}0.0034 \pm 0.0055$ & 
        $\phantom{-}0.0034 \pm 0.0016$ & Lumi \\
    	$\eta_{\text{CP}}\mathcal{A}_{\text{mix}}^{\text{CP}}(B_s^0\to J/\psi \phi)$ & 
        $-0.061 \pm 0.014$ & This & 
        $-0.061\phantom{0} \pm 0.011$ & 
        $-0.0612\phantom{0} \pm 0.0039$ & \cite{LHCb:2018roe}\\
	    \midrule
	    $\phantom{\eta_{\text{CP}}}\mathcal{A}_{\text{dir}}^{\text{CP}}(B_d^0\to J/\psi \rho^0)$ & 
        $-0.063 \pm 0.056 \pm 0.019$ & \cite{LHCb:2014xpr} & 
        $-0.067 \pm 0.014 \pm 0.019$ & 
        $-0.067 \pm 0.004 \pm 0.019$ & Lumi \\
    	$\eta_{\text{CP}}\mathcal{A}_{\text{mix}}^{\text{CP}}(B_d^0\to J/\psi \rho^0)$ & 
        $\phantom{-}0.66\phantom{0} \pm 0.13\phantom{0} \pm 0.09\phantom{0}$ & \cite{LHCb:2014xpr} & 
        $\phantom{-}0.663 \pm 0.034 \pm 0.09\phantom{0}$ & 
        $\phantom{-}0.663 \pm 0.009 \pm 0.09\phantom{0}$ & Lumi\\
	    $\rho$ & $+0.01$ & \cite{LHCb:2014xpr} & $+0.01$ & $+0.01$ & Copy\\
	    \midrule
    	$\phantom{\eta_{\text{CP}}}\mathcal{A}_{\text{dir}}^{\text{CP}}(B_s^0\to D_s^+D_s^-)$ & 
        $0.09 \pm 0.12$ & This & 
        $\phantom{-}0.0001 \pm 0.055$ & 
        $\phantom{-}0.0001 \pm 0.026$ & Lumi \\
    	$\eta_{\text{CP}}\mathcal{A}_{\text{mix}}^{\text{CP}}(B_s^0\to D_s^+D_s^-)$ & 
        $0.02 \pm 0.10$ & This & 
        $-0.064\phantom{1} \pm 0.035$ & 
        $-0.064\phantom{1} \pm 0.009$ & \cite{LHCb:2018roe}\\
	    \midrule
    	$\phantom{\eta_{\text{CP}}}\mathcal{A}_{\text{dir}}^{\text{CP}}(B_d^0\to D^+D^-)$ & 
        $-0.004 \pm 0.078$ & This & 
        $-0.002 \pm 0.044$ & 
        $-0.002 \pm 0.021$ & Lumi \\
    	$\eta_{\text{CP}}\mathcal{A}_{\text{mix}}^{\text{CP}}(B_d^0\to D^+D^-)$ & 
        $\phantom{-}0.706 \pm 0.070$ & This & 
        $\phantom{-}0.706 \pm 0.058$ & 
        $\phantom{-}0.706 \pm 0.043$ & Lumi\\
        \bottomrule
    \end{tabular}
    \caption{Overview of the experimental inputs used in this paper, and their expected improvements by summer 2026 (LHCb 23 fb$^{-1}$ + Belle-II 1 ab$^{-1}$), and by the end of the HL-LHC programme (LHCb 300 fb$^{-1}$), which coincides with the end of Belle-II programme (50 ab$^{-1}$).}
    \label{tab:inputs_future}
\end{sidewaystable}

\begin{figure}
    \centering
    \includegraphics[width=0.49\textwidth]{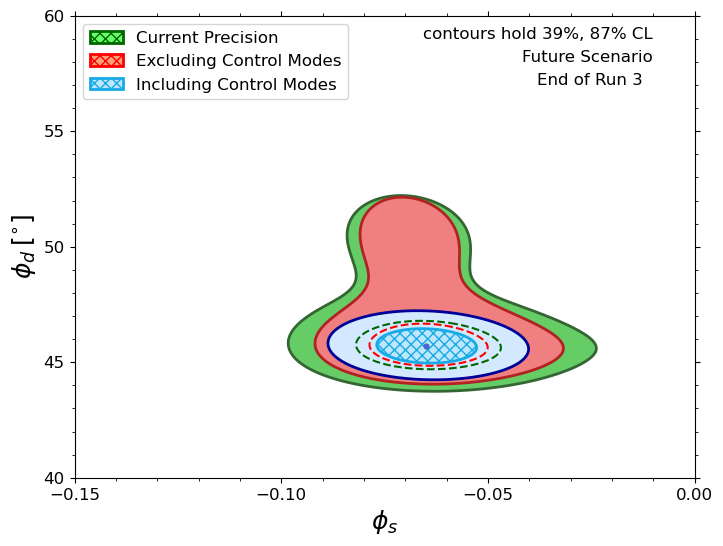}
    \includegraphics[width=0.49\textwidth]{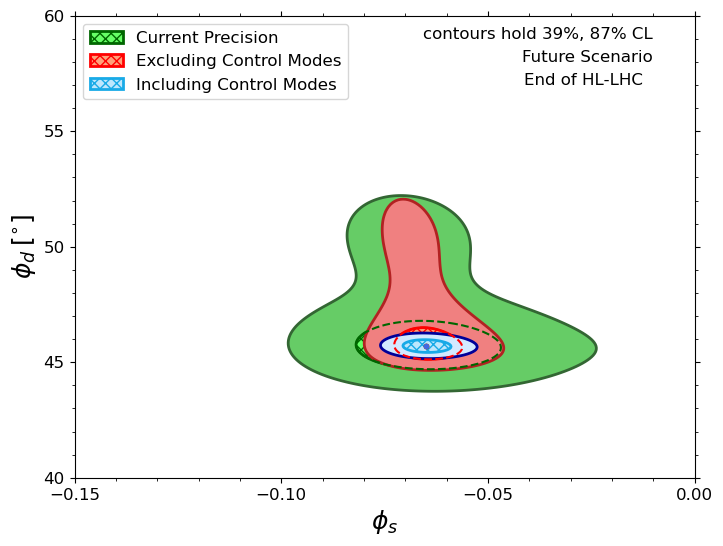}
    \caption{Two-dimensional confidence regions of the nominal fit for $\phi_d$ and $\phi_s$ from the CP asymmetries in the $B\to J/\psi X$ and $B\to DD$ decays.
    Left: Comparison between the \emph{Incl.} and \emph{Excl.\ Penguin} scenarios for the expected situation after summer 2026;
    Right: Comparison between the \emph{Incl.} and \emph{Excl.\ Penguin} scenarios for the expected situation after the end of the HL-LHC programme.
    }
    \label{fig:Future_phi2}
\end{figure}

\begin{figure}
    \centering
    \includegraphics[width=0.49\textwidth]{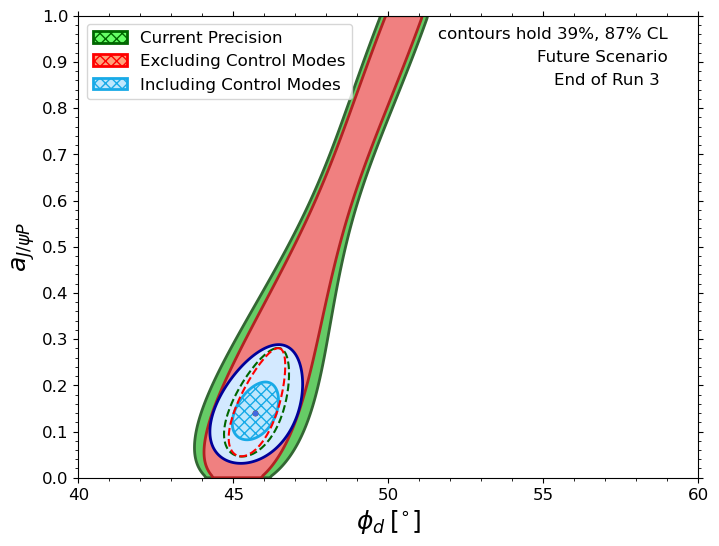}
    \includegraphics[width=0.49\textwidth]{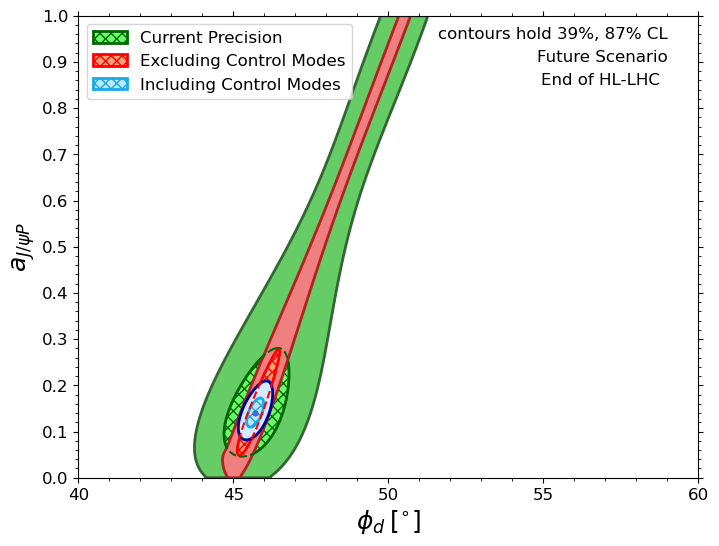}
    
    \includegraphics[width=0.49\textwidth]{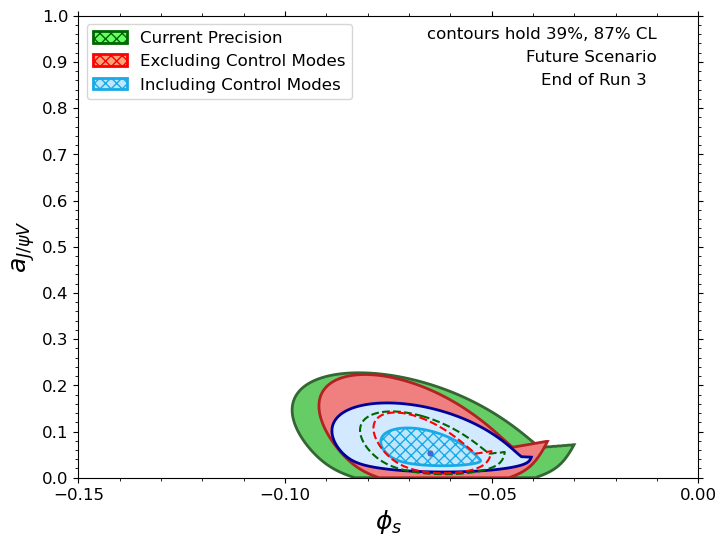}
    \includegraphics[width=0.49\textwidth]{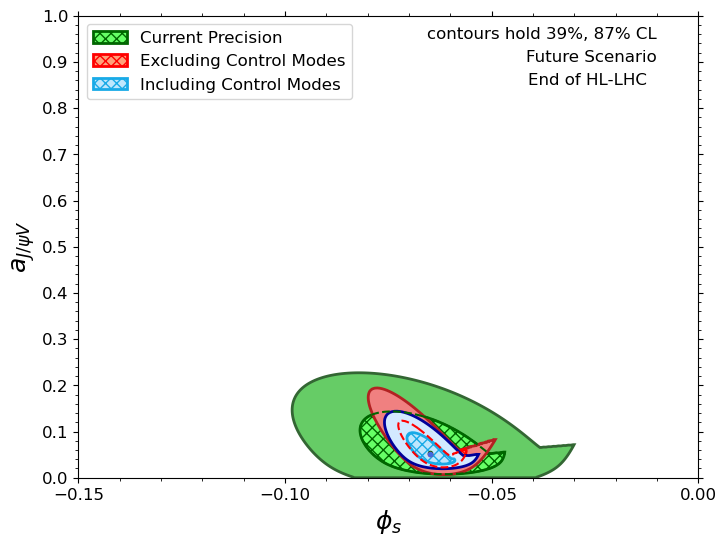}

    \includegraphics[width=0.49\textwidth]{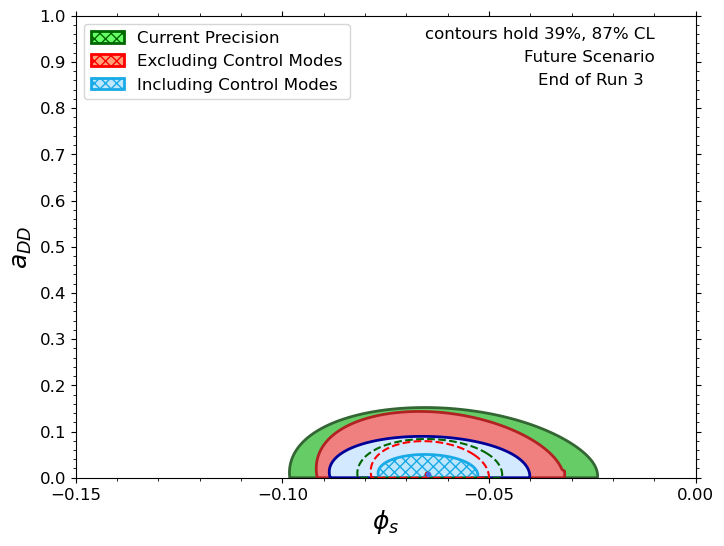}
    \includegraphics[width=0.49\textwidth]{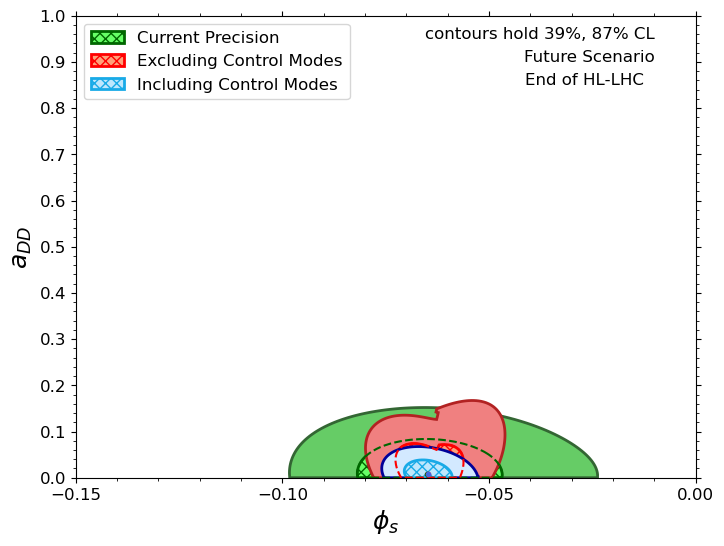}
    \caption{Two-dimensional confidence regions of the nominal fit for $\phi_d$ and $\phi_s$ from the CP asymmetries in the $B\to J/\psi X$ and $B\to DD$ decays.
    Left: Comparison between the \emph{Incl.} and \emph{Excl.\ Penguin} scenarios for the expected situation after summer 2026;
    Right: Comparison between the \emph{Incl.} and \emph{Excl.\ Penguin} scenarios for the expected situation after the end of the HL-LHC programme.
    }
    \label{fig:Future_phi}
\end{figure}

\begin{figure}
    \centering
    \includegraphics[width=0.49\textwidth]{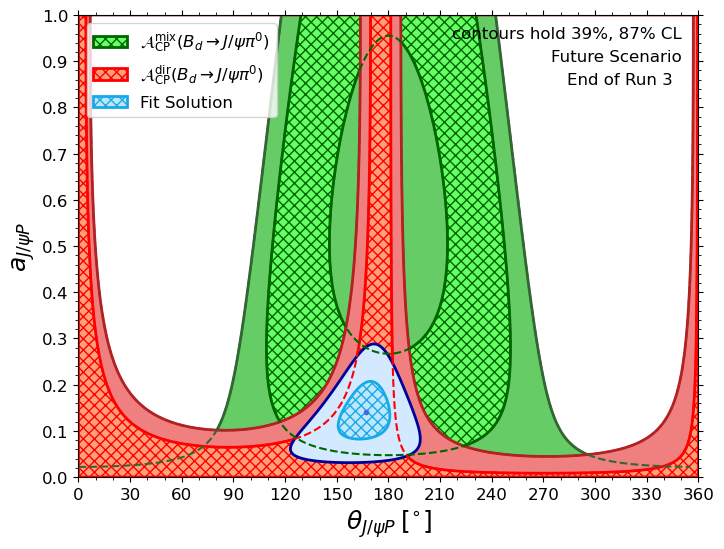}
    \includegraphics[width=0.49\textwidth]{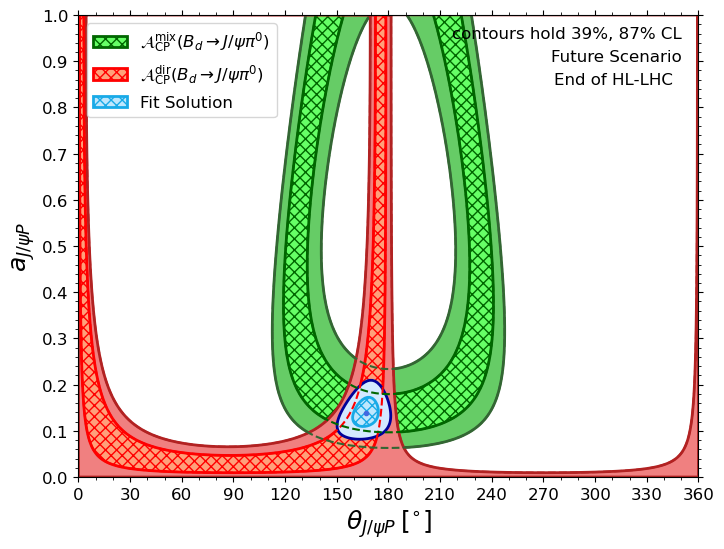}
    
    \includegraphics[width=0.49\textwidth]{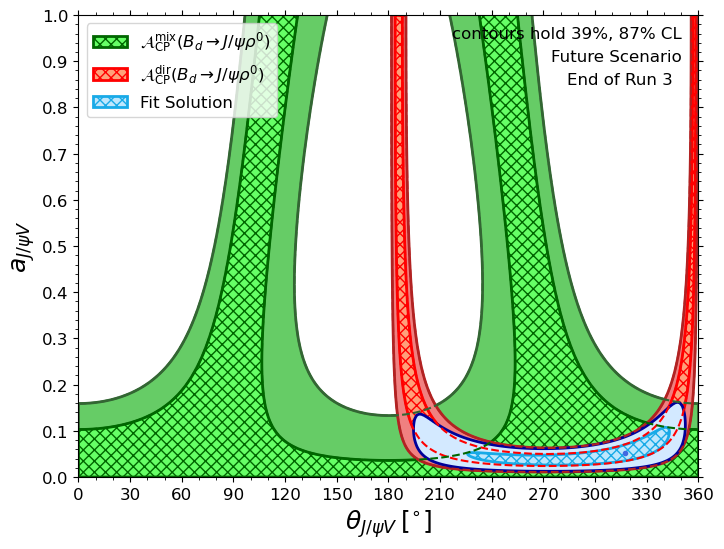}
    \includegraphics[width=0.49\textwidth]{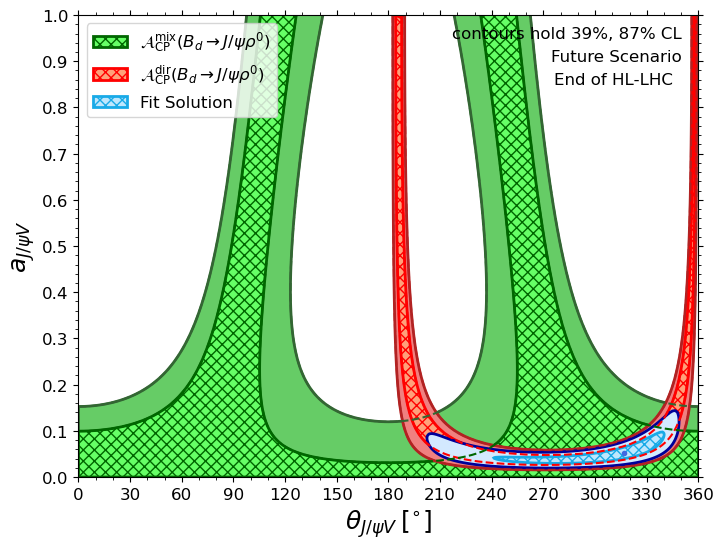}

    \includegraphics[width=0.49\textwidth]{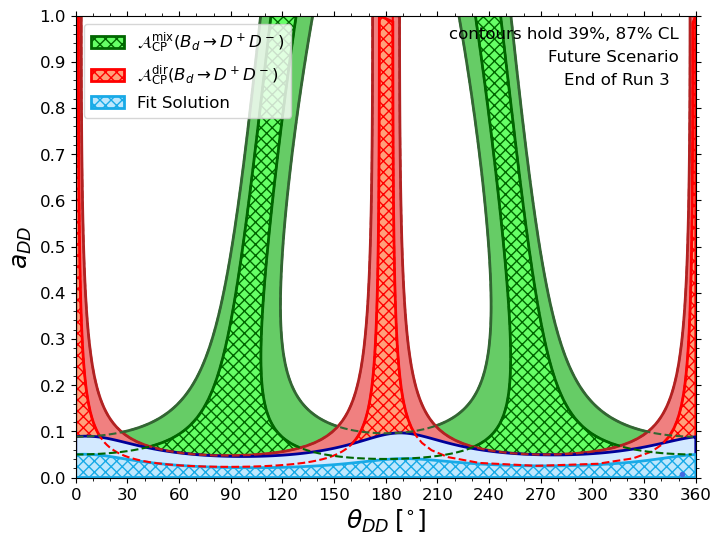}
    \includegraphics[width=0.49\textwidth]{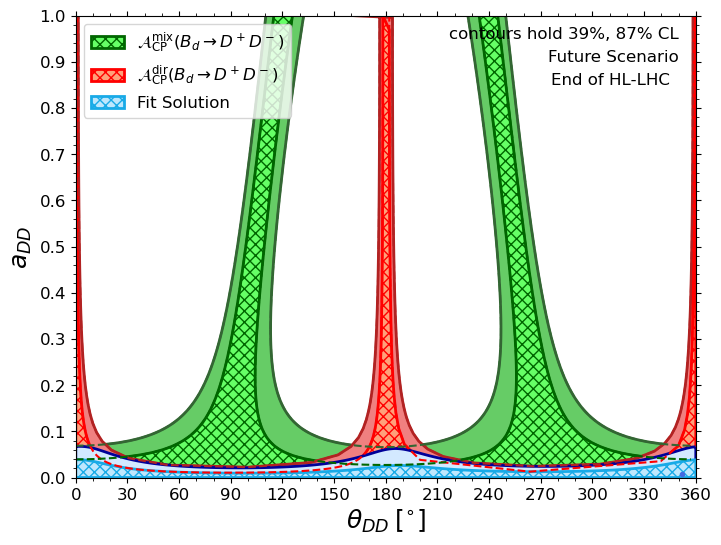}
    \caption{Two-dimensional confidence regions of the nominal fit for the penguin parameters from the CP asymmetries in the $B\to J/\psi X$ and $B\to DD$ decays.
    Left: Incl.\ Penguins end of Run 3 scenario;
    Right: Incl.\ Penguins end of HL-LHC scenario.
    Note that the contours for $\mathcal{A}_{\text{CP}}^{\text{dir}}$ and $\mathcal{A}_{\text{CP}}^{\text{mix}}$ are added for illustration only.
    They include the best fit solutions for $\phi_d$, $\phi_s$ and $\gamma$ as Gaussian constraints.
    }
    \label{fig:Future_theta}
\end{figure}

\paragraph{Fit Solutions}
The numerical results from the four future scenarios are compared to the nominal fit using the current experimental precisions in Table \ref{tab:compare_future}.
Note that these differ slightly from the results in Eqs.\ \eqref{eq:nominal_fit_I}--\eqref{eq:nominal_fit_IV} due to the applied shifts to the central values of the CP asymmetries.
The two-dimensional confidence regions for $\phi_s$ versus $\phi_d$ obtained from the \emph{Incl.} and \emph{Excl.\ Penguin} scenarios are compared to the current precision in Fig.\ \ref{fig:Future_phi2}, while the confidence regions for $\phi_q$ versus the penguin parameters $a$ are shown in Fig.\ \ref{fig:Future_phi}.
For the two \emph{Incl.\ Penguins} scenarios, the two-dimensional confidence regions for the three sets of penguin parameters are shown in Fig.\ \ref{fig:Future_theta}.

\begin{table}
    \resizebox{\textwidth}{!}{
    \renewcommand{\baselinestretch}{1.25}
    \selectfont
    \centering
    \begin{tabular}{|c|c|c|c|c|c|}
        \toprule
        & Current & End of Run 3 & End of Run 3 & End HL-LHC & End HL-LHC \\
        Obs & Precision & {\small Excl.\ Penguins} & {\small Incl.\ Penguins} & {\small Excl.\ Penguins} & {\small Incl.\ Penguins} \\
        \midrule
$\phi_s$ &
$-0.065_{-0.017}^{+0.018}$ & $-0.065_{-0.014}^{+0.015}$ & $-0.065 \pm 0.012$ & $-0.0649_{-0.0077}^{+0.0087}$ & $-0.0649_{-0.0056}^{+0.0060}$ \\
$\phi_s \:[^{\circ}]$ & 
$-3.72_{-0.97}^{+1.03}$ & $-3.72_{-0.80}^{+0.86}$ & $-3.72 \pm 0.69$ & $-3.72_{-0.44}^{+0.50}$ & $-3.72_{-0.32}^{+0.34}$ 
\\
$\phi_d \:[^{\circ}]$ &
$45.7_{-1.0}^{+1.1}$ & $45.69_{-0.85}^{+0.98}$ & $45.70_{-0.74}^{+0.76}$ & $45.69_{-0.57}^{+0.81}$ & $45.70_{-0.29}^{+0.34}$ \\
$a_{J/\psi P}$ &
$0.14_{-0.09}^{+0.14}$ & $0.14_{-0.09}^{+0.14}$ & $0.140_{-0.058}^{+0.067}$ & $0.14_{-0.091}^{+0.14}$ & $0.140_{-0.030}^{+0.033}$ \\
$\theta_{J/\psi P} \:[^{\circ}]$ &
$167_{-32}^{+20}$ & $167_{-31}^{+18}$ & $167_{-16}^{+14}$ & $167_{-29}^{+12}$ & $167.0_{-8.0}^{+7.2}$ \\
$a_{J/\psi V}$ &
$0.053_{-0.045}^{+0.091}$ & $0.053_{-0.043}^{+0.088}$ & $0.053_{-0.027}^{+0.055}$ & $0.052_{-0.030}^{+0.071}$ & $0.052_{-0.022}^{+0.046}$ \\
$\theta_{J/\psi V} \:[^{\circ}]$ &
$317_{-120}^{+38}$ & $317_{-118}^{+36}$ & $317_{-91}^{+26}$ & $317_{-104}^{+29}$ & $317_{-76}^{+23}$ \\
$a_{DD}$ &
$0.008_{-0.008}^{+0.081}$ & $0.008_{-0.008}^{+0.073}$ & $0.008_{-0.008}^{+0.043}$ & $0.008_{-0.008}^{+0.066}$ & $0.008_{-0.008}^{+0.031}$ \\
$\theta_{DD} \:[^{\circ}]$ &
$352_{-352}^{+8}$ & $352_{-352}^{+8}$ & $352_{-352}^{+8}$ & $352_{-352}^{+8}$ & $352_{-352}^{+8}$ \\
        \bottomrule
    \end{tabular}
    }
    \caption{Comparison of the solutions from the nominal fit scenario between the current precision and our four future scenarios.}
    \label{tab:compare_future}
\end{table}


Compared to today, the \emph{Excl.\ Penguin} scenarios will improve the determination of $\phi_d$ by approximately 15\% by \emph{End of Run 3} and approximately 25\% by the \emph{End of HL-LHC}.
In addition, the precision on $\phi_d$ is 15\% (Run 3) to 50\% (HL-LHC) better in the \emph{Incl.\ Penguins} scenarios compared to the \emph{Excl.\ Penguins} scenarios.
This effect is even larger when considering the 87\% confidence region, due to the strong correlation between $\phi_d$ and $a_{J/\psi P}$.
There is thus a significant gain in precision by remeasuring the CP asymmetries in the penguin control modes with the complete data sets.

Regarding the determination of $\phi_s$, the increased luminosity provided by the HL-LHC will reduce the uncertainty by at least 50\% compared to today.
The precision on $\phi_s$ is 15\% (Run 3) to 30\% (HL-LHC) better in the \emph{Incl.\ Penguins} scenarios compared to the \emph{Excl.\ Penguins} scenarios.
Also for this observable significant gains can thus be made by remeasuring the CP asymmetries in the penguin control modes with the complete data sets, even though the correlation between $\phi_s$ and the penguin parameters is much smaller than for $\phi_d$.
Controlling the penguin contributions is thus critical to maximally benefit from the Belle-II and HL-LHC programmes in the search for BSM physics.

\section{Conclusions}\label{sec:conclusion}
The mixing-induced CP asymmetries in the decays $B_d^0\to J/\psi K^0$, $B_s^0\to J/\psi\phi$ and $B_s^0\to D_s^+D_s^-$ are key objectives of the LHC and Belle-II physics programmes because they allow for a high-precision determination of the $B_q^0$--$\bar B_q^0$ mixing phases $\phi_d$ and $\phi_s$.
However, the experimental measurements do not directly determine these two phases, but instead give access to effective mixing phases $\phi_q^{\text{eff}}$.
The effective phases differ from $\phi_d$ and $\phi_s$ due to decay-channel-specific hadronic phase shifts $\Delta \phi_q$ caused by the presence of doubly Cabibbo-suppressed penguin topologies in the decay processes.
In this paper, we used the $SU(3)$ flavour symmetry of QCD to relate the subleading effects from these penguin topologies to counterparts in suitably chosen control modes.
For our nominal fit scenario, we use experimental inputs from the following control channels: $B_s^0\to J/\psi K_{\text{S}}^0$, $B_d^0\to J/\psi\pi^0$, $B_d^0\to J/\psi\rho^0$ and $B_d^0\to D^+D^-$.
In particular, our results include the recently presented updated measurements from LHCb for the effective mixing phase $\phi_s^{\text{eff}}$ in $B_s^0\to D_s^+D_s^-$ \cite{LHCb:2024gkk} and the CP asymmetries in $B_d^0\to D^+D^-$ \cite{LHCb:2024gkk}, as well as the new measurement of the CP asymmetries in $B_d^0\to J/\psi\pi^0$ \cite{Belle-II:2024hqw} released by Belle-II.
The extended fit scenario also makes use of the new LHCb measurement of the direct CP asymmetry difference between $B^+\to J/\psi \pi^+$ and $B^+\to J/\psi K^+$ \cite{LHCb:2024exp}. 

Because the mixing-induced CP asymmetries of the control modes depend on $\phi_d$ and $\phi_s$, which we aim to determine with the highest possible precision, we presented a simultaneous analysis of all decay channels.
For the nominal fit scenario, this results in the following values for the mixing phases:
\begin{equation}
    \phi_d = \left(45.7_{-1.0}^{+1.1}\right)^{\circ}\:,
    \qquad
    \phi_s = -0.065_{-0.017}^{+0.018} 
    = \left(-3.72_{-0.97}^{+1.03}\right)^{\circ}
    \:. \label{eq:phi_con}
\end{equation}
As was shown in Fig.\ \ref{fig:phi_SU3_scan}, these results are robust against potential non-factorisable $SU(3)$-breaking effects.

Comparing the results in Eq.\ \eqref{eq:phi_con} above with the effective mixing phases in Eqs.\ \eqref{eq:phis_eff} and \eqref{eq:phid_eff}, respectively, we observe that the hadronic phase shifts $\Delta\phi_q$ turn out to be small.
This is a highly non-trivial conclusion following from the experimental data that can only be made because we have performed a detailed analysis of the penguin contributions in $B_d^0\to J/\psi K^0$, $B_s^0\to J/\psi\phi$ and $B_s^0\to D_s^+D_s^-$.
In principle, the impact of the penguin topologies could have been much larger.
Although the picture emerging from the current data is favourable, and the uncertainties associated with the penguin effects are at the moment smaller than the uncertainties from the experimental inputs, this situation is expected to change over the next 10 years.
Therefore, in order to reach a precision on $\phi_s$ below 10 mrad and/or on $\phi_d$ below $1^{\circ}$, as forecasted for the end of HL-LHC and Belle-II physics programmes, we will need to further improve our understanding of these penguin effects.
It is thus of the utmost importance that also the CP asymmetries of the control modes get measured with the highest possible precision.
In particular, we look forward to an updated measurement of the $B_s^0\to J/\psi K_{\text{S}}^0$ CP asymmetries.
From a theoretical point of view, this decay is the best possible partner of the golden mode $B_d^0\to J/\psi K^0$, because of the one-to-one relation between all their decay topologies, thereby representing an ideal $U$-spin pair.
Regarding the $B_s^0\to J/\psi\phi$ decay, we advocate for polarisation-dependent measurements of the CP asymmetries (or effective mixing phase), as the hadronic phase shifts can depend on the polarisation of the final state.
With these improved measurements and the resulting improved understanding of the penguin contributions, we are looking forward to continuing the testing of the SM with higher and higher precision through the $B_q^0$--$\bar B_q^0$ mixing phases $\phi_d$ and $\phi_s$ for many years to come.

\section*{Acknowledgements}
This research has been supported by the Netherlands Organisation for Scientific Research (NWO), and the Deutsche Forschungsgemeinschaft (DFG, German Research Foundation) under grant 396021762 - TRR 257.

\phantomsection 
\addcontentsline{toc}{section}{References}
\setboolean{inbibliography}{true}
\bibliographystyle{LHCb}
\bibliography{references}

@article{ATLAS:2025lrr,
    author = "",
    collaboration = "ATLAS, Belle II, CMS and LHCb Collaborations",
    title = "{Projections for key measurements in heavy flavour physics}",
    eprint = "2503.24346",
    archivePrefix = "arXiv",
    primaryClass = "hep-ex",
    reportNumber = "ATL-PHYS-PUB-2025-020, CMS-BPH-25-001, LHCb-PUB-2025-008, Belle II
  Preprint 2025-007, KEK Preprint 2025-5",
    month = "3",
    year = "2025"
}

@article{Barel:2020jvf,
    author = "Barel, Marten Z. and De Bruyn, Kristof and Fleischer, Robert and Malami, Eleftheria",
    title = "{In pursuit of new physics with $B_d^0\to J/\psi K^0$ and $B_s^0\to J/\psi\phi$ decays at the high-precision frontier}",
    eprint = "2010.14423",
    archivePrefix = "arXiv",
    primaryClass = "hep-ph",
    reportNumber = "Nikhef-2020-036",
    doi = "10.1088/1361-6471/abf2a2",
    journal = "J. Phys. G",
    volume = "48",
    number = "6",
    pages = "065002",
    year = "2021"
}

@article{Bel:2015wha,
    author = "Bel, Lennaert and De Bruyn, Kristof and Fleischer, Robert and Mulder, Mick and Tuning, Niels",
    title = "{Anatomy of $ B\to D\overline{D} $ decays}",
    eprint = "1505.01361",
    archivePrefix = "arXiv",
    primaryClass = "hep-ph",
    reportNumber = "NIKHEF-2015-019",
    doi = "10.1007/JHEP07(2015)108",
    journal = "JHEP",
    volume = "07",
    pages = "108",
    year = "2015"
}

@article{Belle-II:2018jsg,
    author = "Altmannshofer, W. and others",
    editor = "Kou, E. and Urquijo, P.",
    collaboration = "Belle II Collaboration",
    title = "{The Belle II Physics Book}",
    eprint = "1808.10567",
    archivePrefix = "arXiv",
    primaryClass = "hep-ex",
    reportNumber = "KEK Preprint 2018-27, BELLE2-PUB-PH-2018-001, FERMILAB-PUB-18-398-T, JLAB-THY-18-2780, INT-PUB-18-047, UWThPh 2018-26",
    doi = "10.1093/ptep/ptz106",
    journal = "PTEP",
    volume = "2019",
    number = "12",
    pages = "123C01",
    year = "2019",
    note = "[Erratum: PTEP 2020, 029201 (2020)]"
}

@article{Belle-II:2024lwr,
    author = "Adachi, I. and others",
    collaboration = "Belle II Collaboration",
    title = "{New graph-neural-network flavor tagger for Belle II and measurement of $
sin 2\phi_1$ in $B^0\to J/\psi K_{\mathrm S}^0$ decays}",
    eprint = "2402.17260",
    archivePrefix = "arXiv",
    primaryClass = "hep-ex",
    reportNumber = "Belle II Preprint 2024-006, KEK Preprint 2023-53",
    doi = "10.1103/PhysRevD.110.012001",
    journal = "Phys. Rev. D",
    volume = "110",
    number = "1",
    pages = "012001",
    year = "2024"
}

@article{Belle-II:2024hqw,
    author = "Adachi, I. and others",
    collaboration = "Belle II Collaboration",
    title = "{Observation of time-dependent CP violation and measurement of the branching fraction of $B^0 \to J/\psi \pi^0$ decays}",
    eprint = "2410.08622",
    archivePrefix = "arXiv",
    primaryClass = "hep-ex",
    reportNumber = "Belle II preprint: 2024-018, KEK preprint: 2024-14",
    doi = "10.1103/PhysRevD.111.012011",
    journal = "Phys. Rev. D",
    volume = "111",
    number = "1",
    pages = "012011",
    year = "2025"
}

@article{Belle-II:2024fgp,
    author = "Adachi, I. and others",
    collaboration = "Belle II Collaboration",
    title = "{Observation of the decay $B^0\to J/\psi\omega$ at Belle II}",
    eprint = "2412.12338",
    archivePrefix = "arXiv",
    primaryClass = "hep-ex",
    reportNumber = "Belle II preprint: 2024-021, KEK preprint: 2024-18",
    doi = "10.1103/PhysRevD.111.032012",
    journal = "Phys. Rev. D",
    volume = "111",
    number = "3",
    pages = "032012",
    year = "2025"
}

@article{Buras:1994ec,
    author = "Buras, Andrzej J. and Lautenbacher, Markus E. and Ostermaier, Gaby",
    title = "{Waiting for the top quark mass, $K^+\to\pi^+\nu\bar\nu$, $B_s^0$--$\bar B_s^0$ mixing and CP asymmetries in $B$ decays}",
    eprint = "hep-ph/9403384",
    archivePrefix = "arXiv",
    reportNumber = "MPI-PH-94-14, TUM-T31-57-94",
    doi = "10.1103/PhysRevD.50.3433",
    journal = "Phys. Rev. D",
    volume = "50",
    pages = "3433--3446",
    year = "1994"
}

@article{Cabibbo:1963yz,
    author = "Cabibbo, Nicola",
    title = "{Unitary symmetry and leptonic decays}",
    doi = "10.1103/PhysRevLett.10.531",
    journal = "Phys. Rev. Lett.",
    volume = "10",
    pages = "531--533",
    year = "1963"
}

@article{Charles:2015gya,
    collaboration = "CKMfitter Group",
    author = "Charles, J. and others",
    title = "{Current status of the Standard Model CKM fit and constraints on $\Delta F=2$ new physics}",
    eprint = "1501.05013",
    archivePrefix = "arXiv",
    primaryClass = "hep-ph",
    reportNumber = "LPT-ORSAY-15-04",
    doi = "10.1103/PhysRevD.91.073007",
    journal = "Phys. Rev. D",
    volume = "91",
    number = "7",
    pages = "073007",
    year = "2015",
    note = "Online updates at \url{ckmfitter.in2p3.fr}"
}

@article{Charles:2017evz,
    author = "Charles, J. and Deschamps, O. and Descotes-Genon, S. and Niess, V.",
    title = "{Isospin analysis of charmless $B$-meson decays}",
    eprint = "1705.02981",
    archivePrefix = "arXiv",
    primaryClass = "hep-ph",
    doi = "10.1140/epjc/s10052-017-5126-9",
    journal = "Eur. Phys. J. C",
    volume = "77",
    number = "8",
    pages = "574",
    year = "2017"
}

@article{Ciuchini:2005mg,
    author = "Ciuchini, M. and Pierini, M. and Silvestrini, L.",
    title = "{The effect of penguins in the $B_d\to J/\psi K^0$ CP asymmetry}",
    eprint = "hep-ph/0507290",
    archivePrefix = "arXiv",
    reportNumber = "RM3-TH-05-5, TUM-HEP-595-05",
    doi = "10.1103/PhysRevLett.95.221804",
    journal = "Phys. Rev. Lett.",
    volume = "95",
    pages = "221804",
    year = "2005"
}

@article{Ciuchini:2011kd,
    author = "Ciuchini, Marco and Pierini, Maurizio and Silvestrini, Luca",
    title = "{Theoretical uncertainty in $\sin 2\beta$: An update}",
    eprint = "1102.0392",
    archivePrefix = "arXiv",
    primaryClass = "hep-ph",
    month = "2",
    year = "2011",
    notes = "{6th International Workshop on the CKM Unitarity Triangle}"
}

@article{CMS:2024znt,
    author = "Hayrapetyan, Aram and others",
    collaboration = "CMS Collaboration",
    title = "{Evidence for $CP$ violation and measurement of $CP$-violating parameters in B$^0_\mathrm{s}$ $\to$ J/$\psi\,\phi$(1020) decays in pp collisions at $\sqrt{s} =$ 13 TeV}",
    eprint = "2412.19952",
    archivePrefix = "arXiv",
    primaryClass = "hep-ex",
    reportNumber = "CMS-BPH-23-004, CERN-EP-2024-300",
    month = "12",
    year = "2024"
}

@article{Davies:2023arm,
    author = "Davies, Jonathan and Jung, Martin and Schacht, Stefan",
    title = "{$\overline{B}\to \overline{D}D $ decays and the extraction of $f_d/f_u$ at hadron colliders}",
    eprint = "2311.16952",
    archivePrefix = "arXiv",
    primaryClass = "hep-ph",
    doi = "10.1007/JHEP01(2024)191",
    journal = "JHEP",
    volume = "01",
    pages = "191",
    year = "2024"
}

@article{DeBruyn:2014oga,
    author = "De Bruyn, Kristof and Fleischer, Robert",
    title = "{A roadmap to control penguin effects in $B^0_d\to J/\psi K_{\rm S}^0$ and $B^0_s\to J/\psi \phi$}",
    eprint = "1412.6834",
    archivePrefix = "arXiv",
    primaryClass = "hep-ph",
    reportNumber = "NIKHEF-2014-031",
    doi = "10.1007/JHEP03(2015)145",
    journal = "JHEP",
    volume = "03",
    pages = "145",
    year = "2015"
}

@article{DeBruyn:2022zhw,
    author = "De Bruyn, Kristof and Fleischer, Robert and Malami, Eleftheria and van Vliet, Philine",
    title = "{New physics in $B_q^0$-$\bar B_q^0$ mixing: present challenges, prospects, and implications for $B_q^0\to\mu^+\mu^-$}",
    eprint = "2208.14910",
    archivePrefix = "arXiv",
    primaryClass = "hep-ph",
    reportNumber = "Nikhef-2022-012",
    doi = "10.1088/1361-6471/acab1d",
    journal = "J. Phys. G",
    volume = "50",
    number = "4",
    pages = "045003",
    year = "2023"
}

@article{DeBruyn:2010hh,
    author = "De Bruyn, Kristof and Fleischer, Robert and Koppenburg, Patrick",
    title = "{Extracting $\gamma$ and penguin topologies through CP violation in $B_s^0\to J/\psi K_{\text{S}}^0$}",
    eprint = "1010.0089",
    archivePrefix = "arXiv",
    primaryClass = "hep-ph",
    reportNumber = "NIKHEF-2010-033",
    doi = "10.1140/epjc/s10052-010-1495-z",
    journal = "Eur. Phys. J. C",
    volume = "70",
    pages = "1025--1035",
    year = "2010"
}

@article{Faller:2008gt,
    author = "Faller, Sven and Fleischer, Robert and Mannel, Thomas",
    title = "{Precision physics with $B_s^0 \to J/\psi \phi$ at the LHC: The quest for new physics}",
    eprint = "0810.4248",
    archivePrefix = "arXiv",
    primaryClass = "hep-ph",
    reportNumber = "CERN-PH-TH-2008-167",
    doi = "10.1103/PhysRevD.79.014005",
    journal = "Phys. Rev. D",
    volume = "79",
    pages = "014005",
    year = "2009"
}

@article{Faller:2008zc,
    author = "Faller, Sven and Jung, Martin and Fleischer, Robert and Mannel, Thomas",
    title = "{The golden modes $B^0\to J/\psi K_{\text{S,L}}$ in the era of precision flavour physics}",
    eprint = "0809.0842",
    archivePrefix = "arXiv",
    primaryClass = "hep-ph",
    reportNumber = "CERN-PH-TH-2008-166, SI-HEP-2008-12",
    doi = "10.1103/PhysRevD.79.014030",
    journal = "Phys. Rev. D",
    volume = "79",
    pages = "014030",
    year = "2009"
}

@article{FLAG:2024oxs,
    author = "Aoki, Y. and others",
    collaboration = "Flavour Lattice Averaging Group",
    title = "{FLAG Review 2024}",
    eprint = "2411.04268",
    archivePrefix = "arXiv",
    primaryClass = "hep-lat",
    reportNumber = "CERN-TH-2024-192, FERMILAB-PUB-24-0785-T",
    doi = "10.1103/nfzp-p5dn",
    journal = "Phys. Rev. D",
    volume = "113",
    number = "1",
    pages = "014508",
    year = "2026",
    note = "Online updates at \url{http://flag.itp.unibe.ch/2024}"
}

@article{Fleischer:1999nz,
    author = "Fleischer, Robert",
    title = "{Extracting $\gamma$ from $B_{s(d)} \to J/\psi K_{\text{S}}$ and $B_{d(s)} \to D_{d(s)}^+ D_{d(s)}^-$}",
    eprint = "hep-ph/9903455",
    archivePrefix = "arXiv",
    reportNumber = "CERN-TH-99-78",
    doi = "10.1007/s100529900099",
    journal = "Eur. Phys. J. C",
    volume = "10",
    pages = "299--306",
    year = "1999"
}

@article{Fleischer:1999sj,
    author = "Fleischer, Robert",
    editor = "Erhan, S. and Krizan, P. and Lohse, T.",
    title = "{Recent theoretical developments in CP violation in the $B$ system}",
    eprint = "hep-ph/9908340",
    archivePrefix = "arXiv",
    reportNumber = "CERN-TH-99-242",
    doi = "10.1016/S0168-9002(00)00003-6",
    journal = "Nucl. Instrum. Meth. A",
    volume = "446",
    pages = "1--17",
    year = "2000"
}

@article{Fleischer:1999zi,
    author = "Fleischer, Robert",
    title = "{Extracting CKM phases from angular distributions of $B_{d,s}$ decays into admixtures of CP eigenstates}",
    eprint = "hep-ph/9903540",
    archivePrefix = "arXiv",
    reportNumber = "CERN-TH-99-92",
    doi = "10.1103/PhysRevD.60.073008",
    journal = "Phys. Rev. D",
    volume = "60",
    pages = "073008",
    year = "1999"
}

@article{Fleischer:2010ca,
    author = "Fleischer, Robert and Serra, Nicola and Tuning, Niels",
    title = "{Tests of Factorization and SU(3) Relations in B Decays into Heavy-Light Final States}",
    eprint = "1012.2784",
    archivePrefix = "arXiv",
    primaryClass = "hep-ph",
    reportNumber = "NIKHEF-2010-048",
    doi = "10.1103/PhysRevD.83.014017",
    journal = "Phys. Rev. D",
    volume = "83",
    pages = "014017",
    year = "2011"
}

@article{Frings:2015eva,
    author = "Frings, Philipp and Nierste, Ulrich and Wiebusch, Martin",
    title = "{Penguin contributions to CP phases in $B_{d,s}$ decays to charmonium}",
    eprint = "1503.00859",
    archivePrefix = "arXiv",
    primaryClass = "hep-ph",
    reportNumber = "TTP15-006",
    doi = "10.1103/PhysRevLett.115.061802",
    journal = "Phys. Rev. Lett.",
    volume = "115",
    number = "6",
    pages = "061802",
    year = "2015"
}

@article{Gronau:2008kk,
    author = "Gronau, Michael and Rosner, Jonathan L.",
    title = "{$B$ decays dominated by $\omega$--$\phi$ mixing}",
    eprint = "0806.3584",
    archivePrefix = "arXiv",
    primaryClass = "hep-ph",
    reportNumber = "EFI-08-19",
    doi = "10.1016/j.physletb.2008.07.016",
    journal = "Phys. Lett. B",
    volume = "666",
    pages = "185--188",
    year = "2008"
}

@article{HFLAV:2022pwe,
    author = "Amhis, Y. and others",
    collaboration = "Heavy Flavour Averaging Group",
    title = "{Averages of $b$-hadron, $c$-hadron, and $\tau$-lepton properties as of 2021}",
    eprint = "2206.07501",
    archivePrefix = "arXiv",
    primaryClass = "hep-ex",
    doi = "10.1103/PhysRevD.107.052008",
    journal = "Phys. Rev. D",
    volume = "107",
    number = "5",
    pages = "052008",
    year = "2023",
    note = "Online update \url{https://hflav.web.cern.ch/}"
}

@article{Jung:2012mp,
    author = "Jung, Martin",
    title = "{Determining weak phases from $B\to J/\psi P$ decays}",
    eprint = "1206.2050",
    archivePrefix = "arXiv",
    primaryClass = "hep-ph",
    reportNumber = "DO-TH-12-15",
    doi = "10.1103/PhysRevD.86.053008",
    journal = "Phys. Rev. D",
    volume = "86",
    pages = "053008",
    year = "2012"
}

@article{Jung:2014jfa,
    author = "Jung, Martin and Schacht, Stefan",
    title = "{Standard model predictions and new physics sensitivity in $B \to DD$ decays}",
    eprint = "1410.8396",
    archivePrefix = "arXiv",
    primaryClass = "hep-ph",
    reportNumber = "DO-TH-13-24, TTP-14-027",
    doi = "10.1103/PhysRevD.91.034027",
    journal = "Phys. Rev. D",
    volume = "91",
    number = "3",
    pages = "034027",
    year = "2015"
}

@article{Kobayashi:1973fv,
    author = "Kobayashi, Makoto and Maskawa, Toshihide",
    title = "{CP violation in the renormalizable theory of weak interaction}",
    reportNumber = "KUNS-242",
    doi = "10.1143/PTP.49.652",
    journal = "Prog. Theor. Phys.",
    volume = "49",
    pages = "652--657",
    year = "1973"
}

@article{LHCb:2014xpr,
    author = "Aaij, Roel and others",
    collaboration = "LHCb Collaboration",
    title = "{Measurement of the CP-violating phase $\beta$ in $B^0\rightarrow J/\psi \pi^+\pi^-$ decays and limits on penguin effects}",
    eprint = "1411.1634",
    archivePrefix = "arXiv",
    primaryClass = "hep-ex",
    reportNumber = "CERN-PH-EP-2014-268, LHCB-PAPER-2014-058",
    doi = "10.1016/j.physletb.2015.01.008",
    journal = "Phys. Lett. B",
    volume = "742",
    pages = "38--49",
    year = "2015"
}

@article{LHCb:2015brj,
    author = "Aaij, Roel and others",
    collaboration = "LHCb Collaboration",
    title = "{Measurement of the time-dependent CP asymmetries in $B_s^0\rightarrow J/\psi K_{\rm S}^0$}",
    eprint = "1503.07055",
    archivePrefix = "arXiv",
    primaryClass = "hep-ex",
    reportNumber = "CERN-PH-EP-2015-064, LHCB-PAPER-2015-005",
    doi = "10.1007/JHEP06(2015)131",
    journal = "JHEP",
    volume = "06",
    pages = "131",
    year = "2015"
}

@article{LHCb:2015esn,
    author = "Aaij, Roel and others",
    collaboration = "LHCb Collaboration",
    title = "{Measurement of CP violation parameters and polarisation fractions in $ {\mathrm{B}}_{\mathrm{s}}^0\to \mathrm{J}/\psi {\overline{\mathrm{K}}}^{\ast 0} $ decays}",
    eprint = "1509.00400",
    archivePrefix = "arXiv",
    primaryClass = "hep-ex",
    reportNumber = "CERN-PH-EP-2015-224, LHCB-PAPER-2015-034",
    doi = "10.1007/JHEP11(2015)082",
    journal = "JHEP",
    volume = "11",
    pages = "082",
    year = "2015"
}

@article{LHCb:2016mag,
    author = "Aaij, Roel and others",
    collaboration = "LHCb Collaboration",
    title = "{Measurement of the CKM angle $\gamma$ from a combination of LHCb results}",
    eprint = "1611.03076",
    archivePrefix = "arXiv",
    primaryClass = "hep-ex",
    reportNumber = "LHCB-PAPER-2016-032, CERN-EP-2016-270",
    doi = "10.1007/JHEP12(2016)087",
    journal = "JHEP",
    volume = "12",
    pages = "087",
    year = "2016",
    note = "The GammaCombo package is available from \url{https://gammacombo.github.io}"
}

@article{LHCb:2018roe,
    author = "Aaij, Roel and others",
    collaboration = "LHCb Collaboration",
    title = "{Physics case for an LHCb Upgrade II - Opportunities in flavour physics, and beyond, in the HL-LHC era}",
    eprint = "1808.08865",
    archivePrefix = "arXiv",
    primaryClass = "hep-ex",
    reportNumber = "LHCB-PUB-2018-009, CERN-LHCC-2018-027",
    month = "8",
    year = "2018"
}

@article{LHCb:2023zcp,
    author = "Aaij, Roel and others",
    collaboration = "LHCb Collaboration",
    title = "{Measurement of CP violation in $B^0\rightarrow\psi(\rightarrow\ell^+\ell^-)K_{\mathrm{S}}^0(\rightarrow\pi^+\pi^-)$ decays}",
    eprint = "2309.09728",
    archivePrefix = "arXiv",
    primaryClass = "hep-ex",
    reportNumber = "LHCb-PAPER-2023-013, CERN-EP-2023-177",
    doi = "10.1103/PhysRevLett.132.021801",
    journal = "Phys. Rev. Lett.",
    volume = "132",
    number = "2",
    pages = "021801",
    year = "2024"
}

@article{LHCb:2024gkk,
    author = "Aaij, Roel and others",
    collaboration = "LHCb Collaboration",
    title = "{Measurement of CP violation in $B^0\to D^+D^-$ and $B_s^0\to D_s^+D_s^-$ decays}",
    eprint = "2409.03009",
    archivePrefix = "arXiv",
    primaryClass = "hep-ex",
    reportNumber = "LHCb-PAPER-2024-027, CERN-EP-2024-217",
    doi = "10.1007/JHEP01(2025)061",
    journal = "JHEP",
    volume = "01",
    pages = "061",
    year = "2025"
}

@article{LHCb:2024exp,
    author = "Aaij, Roel and others",
    collaboration = "LHCb Collaboration",
    title = "{First evidence for direct CP violation in beauty to charmonium decays}",
    eprint = "2411.12178",
    archivePrefix = "arXiv",
    primaryClass = "hep-ex",
    reportNumber = "LHCb-PAPER-2024-031 CERN-EP-2024-286, LHCb-PAPER-2024-031, CERN-EP-2024-286",
    doi = "10.1103/PhysRevLett.134.101801",
    journal = "Phys. Rev. Lett.",
    volume = "134",
    number = "10",
    pages = "101801",
    year = "2025"
}

@article{LHCb:2024xyw,
    author = "Aaij, Roel and others",
    collaboration = "LHCb",
    title = "{Measurement of CP asymmetry in $ {\textrm{B}}_{\textrm{s}}^0\to {\textrm{D}}_{\textrm{s}}^{\mp }{\textrm{K}}^{\pm } $ decays}",
    eprint = "2412.14074",
    archivePrefix = "arXiv",
    primaryClass = "hep-ex",
    reportNumber = "LHCb-PAPER-2024-020, CERN-EP-2024-219",
    doi = "10.1007/JHEP03(2025)139",
    journal = "JHEP",
    volume = "03",
    pages = "139",
    year = "2025"
}

@article{Liu:2013nea,
    author = "Liu, Xin and Wang, Wei and Xie, Yuehong",
    title = "{Penguin pollution in $B\to J/\psi V$ decays and impact on the extraction of the $B_s$--$\bar B_s$ mixing phase}",
    eprint = "1309.0313",
    archivePrefix = "arXiv",
    primaryClass = "hep-ph",
    doi = "10.1103/PhysRevD.89.094010",
    journal = "Phys. Rev. D",
    volume = "89",
    number = "9",
    pages = "094010",
    year = "2014"
}

@article{PDG:2024cfk,
    author = "Navas, S. and others",
    collaboration = "Particle Data Group",
    title = "{Review of particle physics}",
    doi = "10.1103/PhysRevD.110.030001",
    journal = "Phys. Rev. D",
    volume = "110",
    number = "3",
    pages = "030001",
    year = "2024",
    note = "Online updates at \url{https://pdg.lbl.gov}"
}

@article{Seng:2022wcw,
    author = "Seng, Chien-Yeah and Galviz, Daniel and Gorchtein, Mikhail and Mei\ss{}ner, Ulf-G.",
    title = "{Complete theory of radiative corrections to $K_{\ell3}$ decays and the $V_{us}$ update}",
    eprint = "2203.05217",
    archivePrefix = "arXiv",
    primaryClass = "hep-ph",
    doi = "10.1007/JHEP07(2022)071",
    journal = "JHEP",
    volume = "07",
    pages = "071",
    year = "2022"
}

@misc{superkekb,
    note = "See \url{https://www-superkekb.kek.jp/Luminosity_projection/}"
}

@article{Wolfenstein:1983yz,
    author = "Wolfenstein, Lincoln",
    title = "{Parametrization of the Kobayashi--Maskawa matrix}",
    reportNumber = "CMU-HEG83-9",
    doi = "10.1103/PhysRevLett.51.1945",
    journal = "Phys. Rev. Lett.",
    volume = "51",
    pages = "1945",
    year = "1983"
}

\end{document}